\documentclass[twocolumn, aps, 10pt, floatfix, superscriptaddress]{revtex4-1}

\usepackage{lmodern}
\usepackage[T1]{fontenc}
\usepackage[utf8]{inputenc}
\usepackage[english]{babel}

\usepackage{amsmath}
\usepackage{amssymb}
\usepackage{mathtools}
\usepackage{bm}
\usepackage{graphicx,color}
\usepackage{hyperref}
\usepackage{booktabs}
\usepackage{epstopdf}
\usepackage{bbold} 

\hypersetup{
  colorlinks   = true, 
  urlcolor     = blue, 
  linkcolor    = blue, 
  citecolor   = red, 
}

\newcommand{\ket}[1]{|#1\rangle}
\newcommand{\bra}[1]{\langle#1|}

\def\CFT{\textsl{\tiny CFT}}
\def\letter{paper } 

\begin{document}

\title{Conformal fields and operator product expansion in critical quantum spin chains}

\author{Yijian Zou}
\email[]{yzou@pitp.ca}
\affiliation{Perimeter Institute for Theoretical Physics, Waterloo ON, N2L 2Y5, Canada}
\affiliation{University of Waterloo, Waterloo ON, N2L 3G1, Canada}
\author{Ashley Milsted}
\affiliation{Perimeter Institute for Theoretical Physics, Waterloo ON, N2L 2Y5, Canada}
\author{Guifre Vidal}
\affiliation{Perimeter Institute for Theoretical Physics, Waterloo ON, N2L 2Y5, Canada}

\date{\today}

\begin{abstract}
We propose a variational method for identifying lattice operators in a critical quantum spin chain with scaling operators in the underlying conformal field theory (CFT). In particular, this allows us to build a lattice version of the primary operators of the CFT, from which we can numerically estimate the operator product expansion coefficients $C_{\alpha\beta\gamma}^{\CFT}$. We demonstrate the approach with the critical Ising quantum spin chain.
\end{abstract}

\maketitle
Critical phenomena \cite{cardy_scaling_1996,sachdev_quantum_1999}, characterized by universal behaviour and scale invariance, are of broad interest in various fields of physics, including statistical mechanics, condensed matter, and quantum fields. 
At a second order phase transition, scale invariance is often enhanced to a larger symmetry group, the conformal group, and the universal, low-energy physics of the critical system is then described by a conformal field theory (CFT) \cite{Polyakov:1970xd,belavin_infinite_1984,friedan_conformal_1984,di_francesco_conformal_2012}, a field theory that is rigidly constrained by conformal invariance. This is particularly manifest in two dimensions, where the entire CFT is specified by a limited set of conformal data: 
its \textit{central charge} $c^{\CFT}$ together with the list of \textit{conformal dimensions} $(h^{\CFT}_\alpha, \bar{h}^{\CFT}_\alpha)$ and \textit{operator product expansion} (OPE) coefficients $C^{\CFT}_{\alpha\beta\gamma}$ for its \textit{primary} operators $\phi^{\CFT}_\alpha$. From the conformal data one can derive the critical exponents of the phase transition as well as the rest of its universal properties.

There has been enormous progress in identifying possible conformal data for 1+1D CFTs, leading for instance to the famous characterization of unitary minimal models \cite{belavin_infinite_1984,friedan_conformal_1984}. However, given a microscopic Hamiltonian $H = \sum_{j=1}^N h(j)$ for a critical quantum spin chain, numerically computing the conformal data of the emergent CFT remains a challenging task. At the core of the problem is the exponential growth of the Hilbert space with the size of the spin chain. Broadly speaking, two possible computational strategies are available. The first one is based on evaluating ground-state two-point and three-point correlators, which directly yield the conformal dimensions and the OPE coefficients, respectively \cite{degli_investigation_2004,tagliacozzo_2008,
xavier_entanglement_2010,evenbly_quantum_2013,stojevic_2015,evenbly_tensor_2015}. 
A second strategy, outlined by Cardy in the 80s \cite{cardy_conformal_1984, cardy_operator_1986}, is based instead on exploiting the CFT operator-state correspondence \cite{di_francesco_conformal_2012}. This correspondence, denoted $\psi_{\alpha}^{\CFT} \sim\ket{\psi_{\alpha}^{\CFT}}$, relates each scaling operator $\psi_{\alpha}^{\CFT}$ of the CFT with a simultaneous eigenvector $\ket{\psi_{\alpha}^{\CFT}}$ of the Hamiltonian and momentum operators $H^{\CFT}$ and $P^{\CFT}$ of the CFT on a circle of length $L$. In particular, 
\begin{eqnarray} \label{eq:energy}
H^{\CFT}\ket{\psi^{\CFT}_\alpha} &=& \frac{2\pi}{L}\left( \Delta^{\CFT}_\alpha - \frac{c^{\CFT}}{12}\right) \ket{\psi^{\CFT}_\alpha}, \\
P^{\CFT}\ket{\psi^{\CFT}_\alpha} &=& \frac{2\pi}{L} s^{\CFT}_\alpha \ket{\psi^{\CFT}_\alpha}, \label{eq:momentum}
\end{eqnarray}
where the energy and momentum of the state $\ket{\psi^{\CFT}_{\alpha}}$ are given in terms of the scaling dimension $\Delta^{\CFT}_\alpha=h^{\CFT}_\alpha+\bar{h}^{\CFT}_\alpha$ and conformal spin $s^{\CFT}_\alpha=h^{\CFT}_\alpha-\bar{h}^{\CFT}_\alpha$ of the scaling operator $\psi_{\alpha}^{\CFT}$. We can then exploit two classic results: (i) As first pointed out by Cardy \cite{cardy_conformal_1984, blote_1986, cardy_logarithmic_1986, cardy_operator_1986,affleck_universal_1986}, the low energy states $\ket{\psi_{\alpha}}$ of a critical quantum spin chain on the circle  
are in one-to-one correspondence with CFT states, $\ket{\psi_{\alpha}} \sim \ket{\psi^{\CFT}_{\alpha}}$, 
and approximately reproduce the spectrum of energies and momenta \eqref{eq:energy}-\eqref{eq:momentum}, from which we can estimate $c^{\CFT}$ and $\Delta_{\alpha}^{\CFT}$ and $s_{\alpha}^{\CFT}$ (or $h_{\alpha}^{\CFT}$ and $\bar{h}_{\alpha}^{\CFT}$); (ii) Following Koo and Saleur \cite{koo_representations_1994, read_associative-algebraic_2007, dubail_conformal_2010, vasseur_puzzle_2012,
gainutdinov_logarithmic_2013, gainutdinov_lattice_2013, bondesan_chiral_2015}, a Fourier expansion of the lattice Hamiltonian term $h(j)$ results in an approximate lattice realization of the CFT Virasoro generators. These can then be used \cite{milsted_extraction_2017} to identify the specific low energy states of the spin chain, denoted $\ket{\phi_{\alpha}}$, that correspond to CFT \textit{primary} states $\ket{\phi^{\CFT}_{\alpha}}$, which are the ones contributing to the conformal data. In addition, in order to significantly reduce finite size errors in the resulting numerical estimates of $c^{\CFT}$ and $h^{\CFT}_{\alpha}, \bar{h}^{\CFT}_{\alpha}$, in Ref. \cite{Zou_conformal_2018} we recently demonstrated the use of periodic uniform matrix product states (puMPS) \cite{rommer_1997,pirvu_matrix_2012} to study critical chains made of up to several hundreds of quantum spins. 
Overall this second strategy, based on the operator-state correspondence, produces more systematic and accurate results than the approaches \cite{degli_investigation_2004,tagliacozzo_2008,
xavier_entanglement_2010,evenbly_quantum_2013,stojevic_2015,evenbly_tensor_2015} based on two-point and three-point correlators. However, its major drawback is that its does not yield the OPE coefficients $C_{\alpha\beta\gamma}^{\CFT}$.

In this \letter we explain how to identify each local lattice operator $\mathcal{O}$, acting on the spin chain, with a corresponding linear combination of CFT scaling operators, 
\begin{equation} \label{eq:expansion}
\mathcal{O} \sim \sum_{\alpha} a_\alpha \psi_{\alpha}^{\CFT} \equiv \mathcal{O}^{\CFT}.
\end{equation}
More specifically, we show how to numerically compute the first few dominant terms in this expansion, corresponding to the CFT operators with the smallest scaling dimensions. As a main application, we then explain how to extract a lattice estimate $C_{\alpha\beta\gamma}$ of the OPE coefficients $C^{\CFT}_{\alpha\beta\gamma}$, by computing the matrix element 
\begin{equation} \label{eq:latticeOPE}
C_{\alpha\beta\gamma} \equiv \left(\frac{2\pi}{N}\right)^{-\Delta_{\alpha}} \bra{\phi_{\beta}} \phi_{\alpha}(0) \ket{\phi_{\gamma}}
\end{equation}
where $\phi_{\alpha}$ is an approximate lattice realization of the CFT primary operator $\phi^{\CFT}_{\alpha}$ and $\ket{\phi_{\beta}}$ and $\ket{\phi_{\gamma}}$ are a pair of primary states of the critical spin chain. In this way we successfully complete Cardy's ambitious program to extract conformal data from a critical lattice Hamiltonian $H$ by exploiting the operator-state correspondence. We demonstrate the approach, valid for any quantum spin chain, by computing the leading terms of the expansion \eqref{eq:expansion} for all one-site and two-site operators of the critical Ising model, see Table \ref{table}, as well as its non-trivial OPE coefficient $C^{\CFT}_{\sigma\sigma\epsilon}$. We also briefly enumerate other future applications of the method.
  
\emph{Exciting the CFT vacuum with local operators.---} We start by reviewing some basic facts. A $1+1$ dimensional CFT can be quantized on a cylinder $S^1 \times \mathbb{R}$, where the compactified dimension represents space, with coordinate $x\in [0,L)$, and the other dimension represents Euclidean time, with coordinate $\tau \in \mathbb{R}$. On the $\tau=0$ circle we build the Hilbert space, spanned by the states $\ket{\psi_{\alpha}^{\CFT}}$ in Eqs. \eqref{eq:energy}-\eqref{eq:momentum}. Let $\mathcal{O}^{\CFT}(x)$ denote a local operator acting on this circle, with Fourier mode decomposition 
\begin{eqnarray} \label{eq:Fourier1a}
\mathcal{O}^{\CFT}(x) &=& \sum_{s\in \mathbb{Z}} \mathcal{O}^{\CFT,s} e^{-is 2\pi x/L}, \\
\mathcal{O}^{\CFT,s} &\equiv& \frac{1}{L} \int_0^L dx\, \mathcal{O}^{\CFT}(x) \, e^{is 2\pi x/L}.\label{eq:Fourier1b}
\end{eqnarray}
Applying Fourier mode $\mathcal{O}^{\CFT,s}$ on the vacuum $\ket{0^{\CFT}}$ results in an eigenstate of $P^{\CFT}$ with momentum $(2\pi/L)s$. When $\mathcal{O}^{\CFT}(x)$ is a primary field $\phi^{\CFT}_\alpha(x)$, we obtain \cite{Supp}
\begin{equation} \label{eq:exact_primary}
\phi^{\CFT,s}_{\alpha}\ket{0^{\CFT}}=\left(\frac{2\pi}{L}\right)^{\Delta^{\CFT}_\alpha}\sum_{m,\bar{m}\geq 0} c^s_{\alpha,(m,\bar{m})}|\phi^{\CFT}_{\alpha,(m,\bar{m})}\rangle,
\end{equation}
where $\phi^{\CFT}_{\alpha,(m,\bar{m})}(x)\equiv \partial^m \bar{\partial}^{\bar{m}}\phi^{\CFT}_\alpha(x)$ denotes a derivative descendant with spin $s_{\alpha}^{\CFT}+m-\bar{m}$, $\partial \equiv (\partial_\tau - i\partial_x)/2$ is the complex derivative (with $\bar{\partial} = (\partial_\tau + i\partial_x)/2$), and 
\begin{equation}
\label{primary_coeff}
 c^s_{\alpha,(m,\bar{m})}\equiv \delta_{m-\bar{m}+s^{\CFT}_\alpha,s}\sqrt{\frac{\Gamma(m+2h^{\CFT}_\alpha)\Gamma(\bar{m}+2\bar{h}^{\CFT}_\alpha)}{m!\bar{m}!\Gamma(2h^{\CFT}_\alpha)\Gamma(2\bar{h}^{\CFT}_\alpha)}}.
\end{equation}
More generally, analogous expressions can be obtained when $\mathcal{O}^{\CFT}$ is not a primary operator. Here we will use the specific case where $\mathcal{O}^{\CFT}$ is a derivative descendant $\phi^{\CFT}_{\alpha,(k,\bar{k})}$ of the primary $\phi^{\CFT}_{\alpha}$, for which one finds \cite{Supp}
\begin{eqnarray}
&&\phi^{\CFT,s}_{\alpha,(k,\bar{k})}\ket{0^{\CFT}}=\left(\frac{2\pi}{L}\right)^{\Delta^{\CFT}_\alpha+k+\bar{k}}  \times \nonumber\\
&&~~~~\sum_{m,\bar{m}\geq 0} (m+h^{\CFT}_\alpha)^k (\bar{m}+\bar{h}^{\CFT}_\alpha)^{\bar{k}} c^s_{\alpha,(m,\bar{m})}|\phi^{\CFT}_{\alpha,(m,\bar{m})}\rangle.~~~ \label{eq:exact_descendant}
\end{eqnarray}
Finally, the OPE coefficients can be extracted from \cite{Supp}
\begin{equation} \label{eq:OPE}
C^{\CFT}_{\alpha\beta\gamma}=\left(\frac{2\pi}{L}\right)^{-\Delta_\alpha}\bra{\phi^{\CFT}_\gamma} \phi^{\CFT}_\alpha(0)\ket{\phi^{\CFT}_{\beta}},
\end{equation}
where $\phi_{\alpha}^{\CFT}$ is a primary operator and $\ket{\phi_{\beta}^{\CFT}}$ and $\ket{\phi_{\gamma}^{\CFT}}$ are a pair of CFT primary states on the circle.
 
\emph{Lattice operators as CFT scaling operators.---} Consider now a critical quantum spin chain and a local operator $\mathcal{O}$, acting on a small number of spins, to which we would like to assign a linear combination of CFT scaling operators as in Eq.~\eqref{eq:expansion}. In practice we will produce an approximate, truncated expansion of the form
\begin{equation} \label{eq:expansion2}
\mathcal{O} ~\stackrel{\tiny \mbox{approx}}{\sim} ~ \sum_{\alpha \in \mathcal{A}} a_{\alpha} \psi_{\alpha}^{\CFT}\equiv \tilde{\mathcal{O}}^{\CFT} 
\end{equation}
using only operators $\psi^{\CFT}_{\alpha}$ in a preselected finite set $\mathcal{A}$. By optimizing the coefficients $a_{\alpha}$ (see below), we hope to obtain a truncated expansion \eqref{eq:expansion2} such that 
\begin{equation}
\bra{\psi_{\beta}} \mathcal{O} \ket{\psi_{\alpha}} = \bra{\psi_{\beta}^{\CFT}}\tilde{\mathcal{O}}^{\CFT}\ket{\psi_{\alpha}^{\CFT}} + O\left( \frac{1}{N^{\Delta_c}}\right),
\end{equation}
where the matrix elements are between low energy states $\ket{\psi_{\alpha}}$ and $\ket{\psi_{\beta}}$, we have equated the size $N$ of the spin chain with the size $L$ of the CFT circle, and $\Delta_c$ is the lowest scaling dimension among operators not included in $\mathcal{A}$. Thus, the accuracy of the expansion should systematically improve (the subleading finite-size corrections be further reduced) by adding more scaling operators in $\mathcal{A}$.

\textit{Constraining $\tilde{\mathcal{O}}^{\CFT}$ without knowing the OPE.---} In analogy with Eqs.~\eqref{eq:Fourier1a}-\eqref{eq:Fourier1b}, we first Fourier expand $\mathcal{O}$,
\begin{equation}
\mathcal{O}(j) = \sum_{s} \mathcal{O}^{s}e^{-is2\pi j/N},~~~~ \mathcal{O}^s \equiv \frac{1}{N} \sum_{j=1}^N \mathcal{O}(j) e^{is2\pi j/N}.
\end{equation}
Given a finite set $\mathcal{B}$ of low energy states $\ket{\psi_{\beta}}$ and a range $\mathcal{S}$ of values $s$, we can numerically evaluate the matrix elements $b_{\beta,s} \equiv \bra{\psi_\beta}\mathcal{O}^{s} \ket{0}$ between the spin chain ground state $\ket{0}$ and state $\ket{\psi_{\beta}}$, for all $\ket{\psi_{\beta}} \in \mathcal{B}$ and $s \in \mathcal{S}$. With the ability to analytically compute $B^{\alpha}_{\beta,s} \equiv \bra{\psi_{\beta}^{\CFT}}\psi_{\alpha}^{\CFT,s} \ket{0^{\CFT}}$ (using e.g.\ Eqs.~\eqref{eq:exact_primary}-\eqref{eq:exact_descendant}), we can also evaluate the corresponding CFT matrix elements
\begin{equation}
\bra{\psi_\beta^{\CFT}}\tilde{\mathcal{O}}^{\CFT,s} \ket{0^{\CFT}} = \sum_{\alpha \in \mathcal{A}} a_{\alpha} \bra{\psi_{\beta}^{\CFT}}\psi_{\alpha}^{\CFT,s} \ket{0^{\CFT}}. \label{eq:CFTevaluate}
\end{equation} 
In this way we can search for the coefficients $a_\alpha$ such that $\bra{\psi_\beta^{\CFT}}\tilde{\mathcal{O}}^{\CFT,s} \ket{0^{\CFT}}$ best approximates $\bra{\psi_\beta}\tilde{\mathcal{O}}^{s} \ket{0}$ for all relevant $s$ and $\beta$, by minimizing the cost function
\begin{eqnarray}\label{eq:cost}
f_N^{\mathcal{O}}(\{a_{\alpha}\}) &\equiv& \sum_{\beta,s} \left| \bra{\psi_\beta}\mathcal{O}^{s} \ket{0} - \bra{\psi_\beta^{\CFT}}\tilde{\mathcal{O}}^{\CFT,s} \ket{0^{\CFT}}\right|^2~~~~ \\
&=& \sum_{\beta,s} \left| b_{\beta,s} - \sum_{\alpha} a_{\alpha} B^{\alpha}_{\beta,s}\right|^2
\end{eqnarray} 
Importantly, $f_N^{\mathcal{O}}(\{a_{\alpha}\})$ depends only on matrix elements involving the vacuum and one excited state (analogous to a CFT \textit{two-point} correlator) and not on matrix elements involving two excited states (analogous to a \textit{three-point} correlator), so that it does not require any knowledge of the OPE coefficients $C_{\alpha\beta\gamma}^{\CFT}$. 

\textit{Optimization.---} The algorithm is divided into parts I and II, involving states and operators, respectively. 

Part I (low energy states): taking the critical Hamiltonian $H = \sum_{j=1}^N h(j)$ of a periodic spin chain as the only input, we compute low energy eigenstates of $H$, using e.g. exact diagonalization for small $N$ and puMPS for larger $N$ \cite{Zou_conformal_2018}. We then use the techniques of Refs. \cite{milsted_extraction_2017,Zou_conformal_2018} to (i) for each low energy state $\ket{\psi_{\alpha}} \sim \ket{\psi_{\alpha}^{\CFT}}$, obtain estimates $\Delta_{\alpha}$ and $s_{\alpha}$ for its scaling dimension and conformal spin; (ii) identify primary states $\ket{\phi_{\alpha}}\sim \ket{\phi^{\CFT}_{\alpha}}$ and their descendants, thus organizing the low energy states into conformal towers. The above tasks involve a large-$N$ extrapolation. At this point we make a judicious choice of sets $\mathcal{A}$, $\mathcal{B}$, and $\mathcal{S}$ (see example below).

Part II (operators): For a fixed system size $N$, we compute the CFT matrix elements $\bra{\psi_{\beta}^{\CFT}}\psi_{\alpha}^{\CFT,s} \ket{0^{\CFT}}$ for all $\psi_\alpha^{\CFT} \in \mathcal{A}$, $\ket{\psi_\beta^{\CFT}} \in \mathcal{B}$ and $s\in\mathcal{S}$ using the corresponding analytical expressions. However, here we employ the previously \textit{estimated} conformal dimensions $h_{\alpha}\equiv (\Delta_{\alpha} + s_{\alpha})/2, \bar{h}_{\alpha}\equiv (\Delta_{\alpha} - s_{\alpha})/2$ instead of their unknown \textit{exact} values $h_{\alpha}^{\CFT}, \bar{h}_{\alpha}^{\CFT}$, so that no previous knowledge of the emergent CFT is required. Then, for each choice of lattice operator $\mathcal{O}$, we compute $\bra{\psi_\beta}\mathcal{O}^{s} \ket{0}$ for all $\psi_{\beta}\in \mathcal{B}$ and $s\in\mathcal{S}$, and use linear least-squares regression to minimize $f_{N}^{\mathcal{O}}(\{a_{\alpha}\})$ in Eq.~\eqref{eq:cost}, resulting in a set of optimal coefficients $a_{\alpha}(N)$. Finally, we repeat the entire calculation for several values of $N$ and extrapolate to $N \rightarrow \infty$, which results in the coefficients $a_{\alpha}$ in  \eqref{eq:expansion2}.

\begin{figure}
  \includegraphics[width=\linewidth]{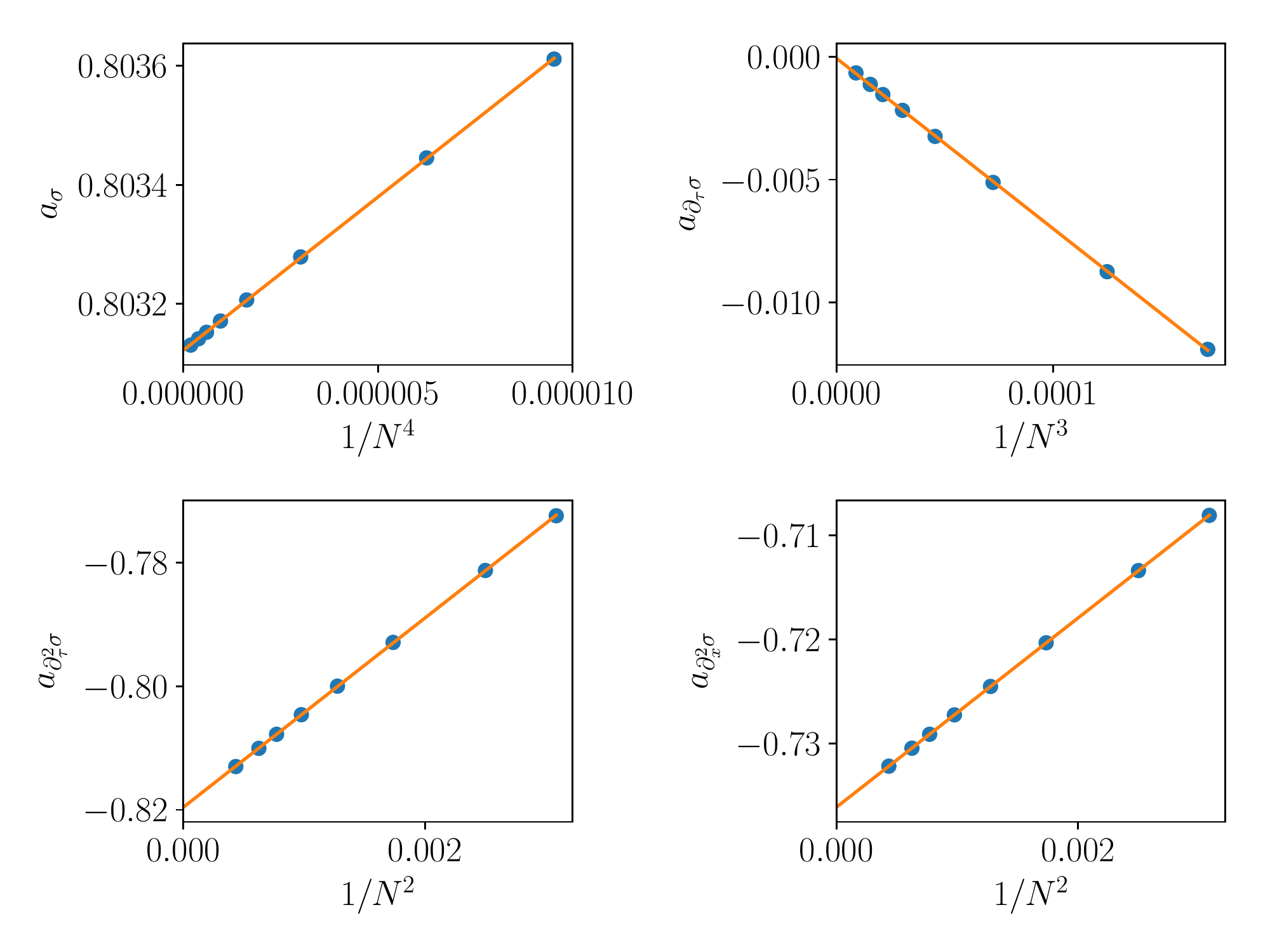}
  \vspace{-2em}
  \caption{Extrapolation of the coefficients of Eq.~\eqref{eq:expansion2} for the lattice operator $\mathcal{O}=XZ+ZX$. The extrapolated values are $a_\sigma = 0.803121$, $a_{\partial_\tau \sigma}=0.0000$, $a_{\partial^2_\tau \sigma}=0.820$, $a_{\partial^2_x \sigma}=-0.736$. See Table \ref{table}.}
  \label{fig:sigma_conv_main}
\end{figure}

\begin{figure}
  \includegraphics[width=1.04\linewidth]{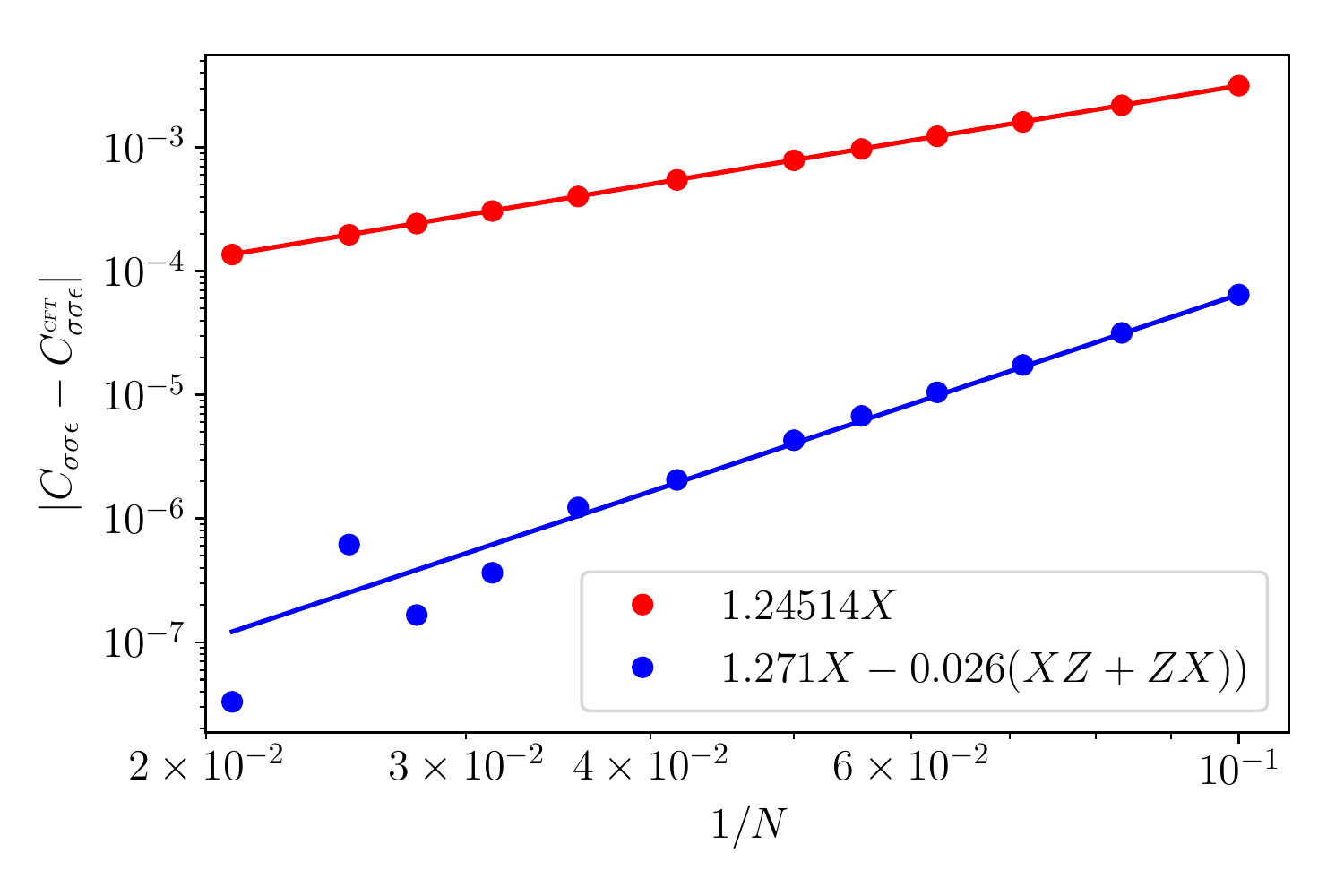}
  \caption{\label{fig:OPE_conv} Error in the OPE coefficient $C_{\sigma\sigma\epsilon}$ as a function of system size for the two lattice realizations \eqref{eq:sigma1} and \eqref{eq:sigma2} of $\sigma^{\CFT}$. Lines correspond to $N^{-2}$ (red) and $N^{-4}$ (blue) scaling.}
\end{figure}

\textit{Example: Critical quantum Ising model.---} To illustrate the approach we consider the critical transverse field Ising Hamiltonian, with local Hamiltonian term $h(j) = -X(j)X(j+1) - Z(j)$. We used puMPS with bond dimension in the range $12 \leq D \leq 22$ to address systems of size $18\leq N \leq 48$, with the largest system size requiring several minutes on a laptop with 4 CPU (2.8 GHz) and 2 GB RAM.
Part I above yields three conformal towers \cite{Zou_conformal_2018}, including the identity (or ground state) tower, present in all CFTs, and two additional conformal towers. We call the corresponding primaries \textit{spin} $\sigma$ and \textit{energy density} $\epsilon$ to follow the standard Ising CFT nomenclature (we reiterate, however, that no previous knowledge of the emergent CFT was used). For the set $\mathcal{A}$ we choose operators
\begin{equation} \label{eq:A1}
\mbox{(identity)} ~\mathbb{1},~~~~~~\mbox{(stress tensor)} ~T, ~\bar{T}~~~~~~
\end{equation}
in the identity conformal tower (notice that $\mathbb{1}$, $T$ and $\bar{T}$ are present in any CFT) and the primaries and first and second derivative descendants in the other two towers, 
\begin{eqnarray} \label{eq:A2}
\mbox{(spin)} ~\sigma, ~~~\partial_\tau \sigma,~ \partial_x \sigma, ~~~  \partial_\tau^2 \sigma, ~\partial_x\partial_\tau \sigma,~ \partial_x^2 \sigma,~~~~~~\\
\mbox{(energy density)}~ \epsilon, ~~~~\partial_\tau \epsilon,~ \partial_x \epsilon, ~~~~  \partial_\tau^2 \epsilon, ~\partial_x\partial_\tau \epsilon,~ \partial_x^2 \epsilon.~~~~~~~ \label{eq:A3}
\end{eqnarray}
For $\mathcal{B}$ we choose all states $\ket{\psi_{\beta}}$ with scaling dimension $\Delta_{\alpha} \leq 3+1/8$, namely the 23 lowest energy states. After normalizing $H$ such that $\Delta_{\mathbb{1}}=0$ and $\Delta_{T}=\Delta_{\bar{T}}=2$ \cite{milsted_extraction_2017}, our estimates for the scaling dimensions of primaries are $\Delta_{\sigma} = 0.1249995$ and $\Delta_{\epsilon} = 0.9999994$, together with exact values $s_{\mathbb{1}} = s_{\sigma} = s_{\epsilon} = 0$, $s_{T}=-s_{\bar{T}}=2$ for the conformal spins. Scaling dimensions and conformal spins of derivative descendants are obtained by adding integers to these values. Finally, for $\mathcal{S}$ we choose $-3 \leq s \leq 3$.

%
\begin{table*}[htbp]
\begin{center}
\footnotesize    
\begin{tabular}{|c|c|c|c|}
\hline
Lattice & CFT & Lattice &CFT \\ \hline

$X$ & $0.803121\sigma-0.017\partial^2_\tau\sigma-0.033\partial^2_x \sigma$ 
& $XY+YX$ & $0.3184i\partial_\tau \epsilon$ 
\\ \hline

$Y$ & $0.8031i\partial_\tau \sigma$ 
& $XY-YX$ & $0.6366(T-\bar{T})$  
\\ \hline

$Z$ & $0.636620\mathbb{1}-0.15915(T+\bar{T})-0.31831\epsilon+0.010\partial^2_\tau\epsilon$
& $XZ+ZX$ & $0.803121\sigma-0.820\partial^2_\tau\sigma-0.736\partial^2_x \sigma$
\\ \hline

$XX$ & $0.636620\mathbb{1}-0.15915(T+\bar{T})+0.31831\epsilon-0.010\partial^2_\tau \epsilon$
& $XZ-ZX$ & $1.205 \partial_x \sigma$
\\ \hline

$YY$ & $-0.212207\mathbb{1}+0.4774(T+\bar{T})+0.31831\epsilon-0.089\partial^2_\tau\epsilon$ 
& $YZ+ZY$ & $2.41 i\partial_\tau \sigma$  
\\ \hline
$ZZ$ & $0.540380\mathbb{1}-0.5403(T+\bar{T})-0.54038\epsilon+0.067\partial^2_\tau\epsilon-0.051\partial^2_x\epsilon$
&   $YZ-ZY$ & $-0.4015i\partial_\tau \partial_x \sigma$ 
\\ \hline

$XIX$ & $0.540380\mathbb{1}-0.5403(T+\bar{T})+0.54038\epsilon-0.067\partial^2_\tau\epsilon+0.051\partial^2_x\epsilon$
& $XZX$ & $0.212207\mathbb{1}-0.4774(T+\bar{T})+0.31831\epsilon-0.089\partial^2_\tau\epsilon$ 
\\ \hline
\end{tabular}
\caption{\label{table} Expansion \eqref{eq:expansion2} for simple lattice operators in the Ising model. Two-spin operators $A(j)B(j+1)$ are organized into terms that are even or odd terms under exchange $j \leftrightarrow j+1$, e.g. $XY \pm YX$. The set $\mathcal{A}$ of CFT operators is given in Eqs. \eqref{eq:A1}-\eqref{eq:A3}. Coefficients smaller than $5\times10^{-3}$ are not shown. The number of significant digits is determined case by case by requiring that a particular digit does not change under extrapolation with different sets of system sizes up to $N=96$  \cite{Supp}. Note that we omit the superscript $\CFT$ on the CFT scaling operators.}
\end{center}
\end{table*}
%

In part II, we evaluated all matrix elements $\bra{\psi^{\CFT}_{\beta}}\psi_{\alpha}^{\CFT} \ket{0^{\CFT}}$ using both Eqs.~\eqref{eq:exact_primary}-\eqref{eq:exact_descendant} and the scaling dimensions and conformal spins quoted above. Then we optimized the truncated expansion \eqref{eq:expansion2} for each Pauli operator $X,Y,Z$ acting on a single spin, for pairs of Pauli operators acting on two continuous spins, etc, see Table~\ref{table} and Fig.~\ref{fig:sigma_conv_main}. Some of the coefficients $a_{\alpha}$ reproduce up to 6 significant digits of their exact value, obtained using the free fermion representation of the Ising model~\cite{Supp}.

\textit{Primary operators and OPE coefficients.---} By inverting the relations in Table \ref{table}, we can build linear combinations of lattice operators whose leading contribution is a targeted CFT scaling operator $\psi^{\CFT}_{\alpha}$. 
For instance, if the target is the spin primary $\sigma^{\CFT}$, its simplest realization is given in terms of the Pauli matrix $X$
\begin{equation} \label{eq:sigma1}
\sigma^{\CFT}~ \stackrel{\tiny \mbox{approx}}{\sim}  ~\mu X,~~~~\mu \approx 1.24514,~~~
\end{equation}
where the approximation can be seen from Table \ref{table} to include both $\partial_x^2 \sigma^{\CFT}$ and $\partial_\tau^2 \sigma^{\CFT}$ as subleading corrections. An improved lattice realization is then given by
\begin{equation} \label{eq:sigma2}
\sigma^{\CFT} ~\stackrel{\tiny \mbox{approx}}{\sim}~ \mu' X + \nu(XZ+ZX),   ~~~
\end{equation}
where $\nu \approx -0.026$, $\mu'\approx 1.27$ and, importantly, the subleading corrections due to $\partial_\tau^2 \sigma^{\CFT}$ have been eliminated This is particularly useful below for the computation of $C_{\sigma \sigma \epsilon}$, seen to be insensitive to the correction $\partial_x^2 \sigma^{\CFT}$ still present in \eqref{eq:sigma2}.
Similarly, for the primary $\epsilon^{\CFT}$ we find
\begin{equation}
~~ \epsilon^{\CFT}~ \stackrel{\tiny \mbox{approx}}{\sim}~ \mu (XX-Z), ~~~~\mu \approx 1.5708, \label{eq:epsilon1}
\end{equation}
with $\partial_{\tau}^2\epsilon^{\CFT}$ as a subleading correction, and the improved
\begin{equation} \label{eq:epsilon2}
~~\epsilon^{\CFT} ~\stackrel{\tiny \mbox{approx}}{\sim}~ \mu'(XX-Z) + \nu(YY+XZX),
\end{equation}
for $\nu\approx -0.19$ and $\mu'\approx 1.76$, with no subleading corrections in $\mathcal{A}$. 
Finally, we can also obtain the stress tensor 
\begin{equation} \label{eq:T}
T^{\CFT},\bar{T}^{\CFT} \stackrel{\tiny \mbox{approx}}{\sim} \mu (XX+Z) \pm \nu (XY-YX) + \nu' \mathbb{1} ,
\end{equation}
where $\mu\approx -1.571$, $\nu \approx 0.7854$, $\nu' \approx 2.000$ and, again, there are no subleading contributions in $\mathcal{A}$. 

Equipped with a lattice realization of the primary operators $\sigma^{\CFT}$ and $\epsilon^{\CFT}$, we can finally use Eq.~\eqref{eq:latticeOPE} (motivated by Eq.~\eqref{eq:OPE}) to compute estimates $C_{\alpha \beta \gamma}$ for the OPE coefficients $C^{\CFT}_{\alpha \beta \gamma}$.  The non-zero coefficients are 
\begin{equation} \label{eq:OPEestimate}
C_{\sigma \sigma \epsilon} \approx 0.500000,~~C_{\sigma \epsilon \sigma} \approx 0.50000,
\end{equation}
obtained by computing $\bra{\sigma} \mathcal{O}_{\sigma}(0)\ket{\epsilon}$ and $\bra{\sigma}\mathcal{O}_{\epsilon}(0)\ket{\sigma}$, with $\mathcal{O}_{\sigma}$ and $\mathcal{O}_{\epsilon}$ the lattice operators in Eqs.\ \eqref{eq:sigma2} and \eqref{eq:epsilon2} and extrapolating to the thermodynamic limit \cite{Supp}. Fig.~\ref{fig:OPE_conv} shows a change in scaling from $N^{-2}$ to $N^{-4}$ in the error of $C_{\sigma\sigma\epsilon}$ when replacing the lattice realization \eqref{eq:sigma1} of $\sigma^{\CFT}$ with the improved lattice realization~\eqref{eq:sigma2}.

\textit{Symmetry, duality, and free fermions.---} The critical Ising model is  invariant under a global spin flip
\begin{equation}
X(j) \rightarrow -X(j), ~~Y(j) \rightarrow -Y(j),~~Z(j) \rightarrow Z(j),
\end{equation}
and (on the real line) under the Kramers-Wannier duality 
\begin{equation}
S_{j} X(j) \rightarrow S_{j} Y(j),~~~~S_{j} Y(j) \rightarrow S_{j+1} X(j+1),
\end{equation}
where $S_{j} \equiv \prod_{i<j}Z(i)$, such that e.g. $X(j)X(j+1) \rightarrow Z(j+1)$ and $Z(j) \rightarrow X(j)X(j+1)$. Our computations above, which made no use of these properties, yield results fully consistent with them, as should be expected. For instance, all CFT and lattice operators in Eqs.~\eqref{eq:sigma1}-\eqref{eq:sigma2} are odd under spin flip symmetry; operators in Eqs.~\eqref{eq:epsilon1}-\eqref{eq:epsilon2} are even under spin flip and odd under the duality transformation; operators in Eq.~\eqref{eq:T} are even under spin flip and duality transformation; finally, the OPE coefficients \eqref{eq:OPEestimate} are the only non-vanishing ones allowed by symmetry. Moreover, the spin model admits a free fermion representation. This allows for an exact computation of some of the coefficients in Table \ref{table} \cite{Supp}, which we used to confirm the accuracy of the numerical results.

\textit{Discussion.---} Given a critical quantum spin chain Hamiltonian $H$ as the only input, in this \letter we have explained how to identify a local lattice operator with a corresponding expansion \eqref{eq:expansion2} in terms of scaling operators $\psi^{\CFT}_{\alpha}$ of the emergent CFT. As demonstrated for the critical Ising model, this allows us to build lattice versions of specific CFT scaling operators. In particular, by targeting primary operators, one can compute OPE coefficients, thereby completing Cardy's program to numerically extract the conformal data from low-energy states of a critical spin chain by exploiting the operator-state correspondence. Our approach, which can be extended to address non-local scaling operators \cite{zou_upcoming_2019-1}, has other useful applications, explored in subsequent work (see also \cite{Supp}). For instance, in a generic critical quantum spin chain one can now modify the original Hamiltonian by adding relevant (or irrelevant) scaling operators on demand. Then, using e.g.\ the techniques demonstrated in Ref.~\cite{Zou_conformal_2018}, one can study, fully non-perturbatively, the renormalization group flow away from (respectively, back to) the initial CFT, along pre-determined directions. Conversely, given a near-critical lattice Hamiltonian, one can tune it closer to criticality by removing relevant perturbations from it.

\textit{Acknowlegments ---} We thank Qi Hu and Martin Ganahl for many useful discussions, as well as Alexander Zamolodchikov for valuable comments. The authors acknowledge support from the Simons Foundation (Many Electron Collaboration) and Compute Canada. Research at Perimeter Institute is supported by the Government of
Canada through the Department of Innovation, Science and Economic Development Canada and by the Province of Ontario through the Ministry of Research, Innovation and Science.

\bibliography{OPE_CFT}

\clearpage

\appendix

\section{General CFT results}
In this section, we derive CFT matrix elements that are used in the main text. While in the main text we have used the superscript ${}^\CFT$ to indicate that a state or operator belong to the CFT, in this section we omit the superscript if it is clear from the context that we are referring to a CFT quantity.

\subsection{CFT matrix elements involving the ground state}
 
We are first concerned with matrix elements involving the ground state and one excited state. They are equivalent to two-point correlations and do not depend on the OPE coefficients of the CFT.

A 1+1 dimensional CFT can be quantized on a cylinder with complex coordinates $w=\tau+ix$ and $\bar{w}=\tau-ix$, where $-\infty<\tau<+\infty$ represents the imaginary time coordinate and $0\leq x <L$ represents the spatial (periodic) coordinate. A primary field $\phi(\tau,x)$ can be mapped to the complex plane with coordinates $(z,\bar{z})$ by the conformal transformation
\begin{equation}
z=e^{2\pi w/L}, 
\end{equation}
which using that $dz/dw = 2\pi z/L$ leads to
\begin{eqnarray}
\label{ExpTransformPrimary}
\phi(\tau,x)&=& \left(\frac{dz}{dw}\right)^{h} \left(\frac{d\bar{z}}{d\bar{w}}\right)^{\bar{h}}\phi(z,\bar{z})\\
&=&\left(\frac{2\pi}{L}\right)^{\Delta} z^h \bar{z}^{\bar{h}}\phi(z,\bar{z}),
\end{eqnarray}
where $h,\bar{h}$ are the conformal dimensions of $\phi(z,\bar{z})$, and $\Delta=h+\bar{h}$ is its scaling dimension.

On the complex plane, the field can be Laurent expanded around the origin,
\begin{equation}
\phi(z,\bar{z})=\sum_{m\in Z-h,\bar{m}\in Z-\bar{h}}z^{-m-h}\bar{z}^{-\bar{m}-\bar{h}}\phi_{m,\bar{m}},
\end{equation}
which becomes Fourier expansion on a time slice of the cylinder,
\begin{equation}
\label{FourierModePrimary}
\phi(0,x)=\left(\frac{2\pi}{L}\right)^{\Delta} \sum_{m,\bar{m}\in Z}e^{2\pi i(m-\bar{m}+s_{\phi})x/L} \phi_{-m-h,-\bar{m}-\bar{h}},
\end{equation}
where $s_\phi=h-\bar{h}$ is the conformal spin.

Applying a Fourier mode on the ground state gives
\begin{equation}
\phi_{-m-h,-\bar{m}-\bar{h}} \ket{0} = \left\{ \begin{array}{cl}
\frac{1}{m!\bar{m}!}L^{m}_{-1}\bar{L}^{\bar{m}}_{-1}\ket{\phi} &  m,\bar{m}\geq 0\\
0 & \mbox{otherwise}
\end{array}\right.
\end{equation}
Therefore, acting with $\phi(0,x)$ on the ground state creates the state $|\phi\rangle$ and all its derivative descendants,
\begin{equation}
L^{m}_{-1}\bar{L}^{\bar{m}}_{-1}|\phi\rangle=C_{m,\bar{m}}|\phi_{(m,\bar{m})}\rangle,
\end{equation}
where $|\phi_{(m,\bar{m})}\rangle$ is a normalized state with conformal dimensions $(m+h,\bar{m}+\bar{h})$ and
\begin{equation}
C_{m,\bar{m}}=\sqrt{m!\bar{m}! \frac{\Gamma(m+2h)\Gamma(\bar{m}+2\bar{h})}{\Gamma(2h)\Gamma(2\bar{h})}}
\end{equation}
is a normalization constant, which can be obtained by successively applying the Virasoro algebra.

We can then obtain the matrix elements of the Fourier mode
\begin{equation}
\label{FourierModeDef}
\phi^s=\frac{1}{L}\int_0^L dx\, \phi(0,x) e^{-2\pi i sx/L}
\end{equation}
by combining Eqs.~\eqref{FourierModePrimary}-\eqref{FourierModeDef}. Adding the superscript ${}^{\CFT}$ and the subscript ${}_\alpha$ to the fields we finally obtain the equation
\begin{equation}
\label{PrimaryCreation}
\phi^{s}_{\alpha}|0\rangle=\left(\frac{2\pi}{L}\right)^{\Delta_\alpha}\sum_{m,\bar{m}\geq 0} c^s_{\alpha,(m,\bar{m})}|\phi^{\CFT}_{\alpha,(m,\bar{m})}\rangle,
\end{equation}
where 
\begin{equation}
\label{defcs}
 c^s_{\alpha,(m,\bar{m})}=\delta_{m-\bar{m}+s_\alpha,s}\sqrt{\frac{\Gamma(m+2h_\alpha)\Gamma(\bar{m}+2\bar{h}_\alpha)}{m!\bar{m}!\Gamma(2h_\alpha)\Gamma(2\bar{h}_\alpha)}},
\end{equation}
as used in the main text.

Let us generalize the above equations to Fourier modes of descendant operators. Note that the transformation rule Eq.~\eqref{ExpTransformPrimary} does not apply. Descendant fields are defined through OPE
\begin{equation}
T(w)\phi(w',\bar{w}')=\sum_{n=-2}^{\infty}(w-w')^n L_{-n-2}\phi(w',\bar{w}'),
\end{equation}
where in particular, for $n=-2$ and $n=-1$ we have singular terms that involve $L_0 \phi=h\phi$ and $L_{-1}\phi=\partial\phi$. For concreteness, we next consider the simplest descendant $L_{-1}\phi$. We can extract it from the OPE by a contour integral
\begin{equation}
\partial\phi(w',\bar{w}')=\frac{1}{2\pi i}\oint_{w'} dw \, T(w)\phi(w',\bar{w}').
\end{equation}
Note that the operator product should be understood as a time-ordered product. We can deform the coutour to two closed circles $\tau_{\pm}=\mathrm{Re} w'\pm\epsilon$ to obtain $\partial\phi(\tau',x')=$
\begin{eqnarray}
\frac{1}{2\pi}\int_0^L \!dx \left(T(\tau^{+},x)\phi(\tau',x')-\phi(\tau',x')T(\tau^{-},x)\right).~~~~~~
\end{eqnarray}
By transforming $T(z)$ onto the cylinder (analogous to Eq.~\eqref{ExpTransformPrimary} with another Schwarzian term proportional to the central charge on the RHS), we obtain
\begin{equation}
\frac{1}{2\pi}\int_0^L dx\, T(\tau,x)=\frac{2\pi}{L}\left(L_0-\frac{c}{24}\right),
\end{equation}
for any $\tau$. Thus we arrive at
\begin{equation}
\label{derivativeL0}
\partial\phi(\tau,x)=\frac{2\pi}{L}[L_0,\phi(\tau,x)].
\end{equation}
Taking the Fourier mode $s$ on both sides, setting $\tau=0$, and acting on the ground state, we obtain an expression similar to Eq.~\eqref{PrimaryCreation} (we also add superscripts and subscripts to compare with the main text):
\begin{widetext}
\begin{equation}
(\partial\phi^{\CFT}_{\alpha})^{s}|0\rangle=\left(\frac{2\pi}{L}\right)^{\Delta^{\CFT}_\alpha+1} 
\sum_{m,\bar{m}\geq 0} (m+h^{\CFT}_\alpha) c^s_{\alpha,(m,\bar{m})}|\phi^{\CFT}_{\alpha,(m,\bar{m})}\rangle,
\end{equation}
and more generally
\begin{equation}
\label{CFTmatrix}
(\partial^k\bar{\partial}^{\bar{k}}\phi^{\CFT}_{\alpha})^{s}|0\rangle=\left(\frac{2\pi}{L}\right)^{\Delta^{\CFT}_\alpha+k+\bar{k}} 
\sum_{m,\bar{m}\geq 0} (m+h^{\CFT}_\alpha)^k (\bar{m}+\bar{h}^{\CFT}_\alpha)^{\bar{k}} c^s_{\alpha,(m,\bar{m})}|\phi^{\CFT}_{\alpha,(m,\bar{m})}\rangle.
\end{equation}
For later use, let us consider a special case $\partial_x \phi_\alpha^{\CFT}\equiv i(\partial-\bar{\partial})\phi_\alpha^{\CFT}$,
\begin{eqnarray}
(\partial_x \phi_\alpha^{\CFT})^{s}|0\rangle&=&\left(\frac{2\pi}{L}\right)^{\Delta^{\CFT}_\alpha+1} 
i\sum_{m,\bar{m}\geq 0} (m-\bar{m}+s^{\CFT}_\alpha) c^s_{\alpha,(m,\bar{m})}|\phi^{\CFT}_{\alpha,(m,\bar{m})}\rangle \\
\label{dxmatrix}
&=&\left(\frac{2\pi}{L}\right)^{\Delta^{\CFT}_\alpha+1} 
i\sum_{m,\bar{m}\geq 0} s c^s_{\alpha,(m,\bar{m})}|\phi^{\CFT}_{\alpha,(m,\bar{m})}\rangle,
\end{eqnarray}
where the first equality follows from linear combing the cases of $k=1,\bar{k}=0$ and $k=0,\bar{k}=1$ in Eq.~\eqref{CFTmatrix} and $s^{\CFT}_\alpha=h^{\CFT}_\alpha-\bar{h}^{\CFT}_{\alpha}$, and the second equality follows from $\delta_{m-\bar{m}+s^{\CFT}_\alpha,s}$ in Eq.~\eqref{defcs}.
\end{widetext}

Generic descendant operators are composite operators of the stress tensors and the primary operators. They can create states (quasi-primary states) which are not derivative descendant states. An analogous formula can still be derived, but is far more complicated. However, the case for the stress tensor scaling operators $T^{\CFT}$ and $\bar{T}^{\CFT}$ (quasi-primary operators in the conformal tower of the identity) can be derived in a simpler way, since their Fourier modes are Virasoro generators,
\begin{eqnarray}
T^{\CFT,s}&=&\left(\frac{2\pi}{L}\right)^2\left(L^{\CFT}_{-s}-\frac{c^{\CFT}}{24}\delta_{s,0}\right) \\
\bar{T}^{\CFT,s}&=&\left(\frac{2\pi}{L}\right)^2\left(\bar{L}^{\CFT}_{s}-\frac{c^{\CFT}}{24}\delta_{s,0}\right)
\end{eqnarray}
Applying the $L^{\CFT}_{-s}, \bar{L}^{\CFT}_{-s}$ to the ground state produces
\begin{eqnarray}
L^{\CFT}_{-s}|0\rangle&=&C_s |\partial^{s-2}T^{\CFT}\rangle \\
\bar{L}^{\CFT}_{-s}|0\rangle&=&C_s |\bar{\partial}^{s-2}\bar{T}^{\CFT}\rangle,
\end{eqnarray}
for $s\geq 2$, where 
\begin{equation}
C_s=\sqrt{\frac{c^{\CFT}}{12}s(s^2-1)}
\end{equation}
is the normalization constant. Combining the equations above gives the matrix elements of $T^{\CFT,s},\bar{T}^{\CFT,s}$.

In the case of the Ising CFT (see next appendix), all descendants of $\sigma^{\CFT}$ and $\epsilon^{\CFT}$ below level $2$ are derivative descendants. Together with the stress tensor operators, these are all the scaling operators included in the truncated set $\mathcal{A}$. Therefore, the above equations are all that is needed for the particular application considered in the main text.

\subsection{OPE coefficients}
On the complex plane, the OPE coefficients are defined by
\begin{equation}
\label{OPE_def}
C^{\CFT}_{\alpha\beta\gamma}=\langle 0|\phi^{\CFT}_\alpha(\infty)\phi^{\CFT}_\beta(1)\phi_\gamma^{\CFT}(0)|0\rangle.
\end{equation}

Transforming $\phi^{\CFT}_\beta$ onto the cylinder with Eq.~\eqref{ExpTransformPrimary}, we have
\begin{equation}
C^{\CFT}_{\alpha\beta\gamma}=\left(\frac{2\pi}{L}\right)^{-\Delta^{\CFT}_\beta}\langle \phi^{\CFT}_\alpha|\phi^{\CFT}_\beta(\tau=0,x=0)|\phi_\gamma^{\CFT}\rangle.
\end{equation}

Since CFT states on the circle $\tau=0$ of the cylinder are eigenstates of the translation $x\rightarrow x+\delta x$, only the Fourier mode of $\phi^{\CFT}_\beta$ with momentum $s_{\alpha}-s_{\gamma}$ contributes to the above equation. Therefore,
\begin{equation}
C^{\CFT}_{\alpha\beta\gamma}=\left(\frac{2\pi}{L}\right)^{-\Delta^{\CFT}_\beta}\langle \phi^{\CFT}_\alpha|\phi^{\CFT,s_\alpha-s_\gamma}_\beta|\phi_\gamma^{\CFT}\rangle.
\end{equation}
Note that generally the OPE coefficient is invariant under even permutations of its three labels (e.g. $C_{\alpha\beta\gamma} = C_{\beta\gamma\alpha}$), and becomes complex conjugated under odd permutations (e.g.  $C_{\alpha\beta\gamma} = C_{\beta\alpha\gamma}^{*}$).

Any other three point function of the same operators is related to the standard form Eq.~\eqref{OPE_def} by some conformal transformation. Therefore, $C^{\CFT}_{\alpha\beta\gamma}$ determines all three point correlation functions. For example,
\begin{eqnarray}
&\,&\langle \phi^{\CFT}_\alpha|(\partial^n_\tau\phi^{\CFT}_\beta)^{s_\alpha-s_\gamma}|\phi_\gamma^{\CFT}\rangle \nonumber\\
\label{taudescendantmatrix}
&=&\left(\frac{2\pi}{L}\right)^{\Delta^{\CFT}_\beta+n}(\Delta^{\CFT}_\alpha-\Delta^{\CFT}_\gamma)^n C^{\CFT}_{\alpha\beta\gamma} \\
&\,&\langle \phi^{\CFT}_\alpha|(\partial^n_x\phi^{\CFT}_\beta)^{s_\alpha-s_\gamma}|\phi_\gamma^{\CFT}\rangle \nonumber\\
\label{xdescendantmatrix}
&=&\left(\frac{2\pi}{L}\right)^{\Delta^{\CFT}_\beta+n}(i(s^{\CFT}_\alpha-s^{\CFT}_\gamma))^n C^{\CFT}_{\alpha\beta\gamma},
\end{eqnarray}
which follows from Eq.~\eqref{derivativeL0} and its antiholomorphic analogue. Matrix elements of general descendants can also be derived in a more complicated way, which we omit here since we will not need them for the particular application in the main text.

\section{The Ising CFT}

The Ising CFT describes the low energy, long distance, universal physics of many critical lattice models, including e.g. the critical Ising quantum spin chain and the quantum critical axial next-to-nearest neighbor Ising quantum spin chain. The Ising CFT is also the first one in the series of unitary minimal models, which can be solved exactly. In this section we shall review some properties of the Ising CFT that are used in this paper. Unless otherwise stated, objects in this section are CFT objects and we omit the superscript ${}^{\CFT}$.

\subsection{Conformal data}

As any other unitary minimal models, this CFT has a finite number of primary operators, resulting in a finite amount of conformal data. Specifically, the central charge is $c=1/2$ and there are three primary fields, namely the identity operator $\mathbb{1}$ (present in any CFT), the spin operator $\sigma$ and the energy density operator $\sigma$. They all have conformal spin $s_{\mathbb{1}}=s_\sigma=s_\epsilon=0$ and their scaling dimensions are $\Delta_{\mathbb{1}}=0$, $\Delta_\sigma=1/8$, and $\Delta_\epsilon=1$, respectively. The only nonzero OPE coefficients (up to permutations of the indices) are $C_{\alpha\beta\mathbb{1}}=\delta_{\alpha\beta}$ and $C_{\sigma\sigma\epsilon}=1/2$. 

\subsection{Null fields and conformal towers}
The Ising CFT has null fields, whose correlation functions are zero and therefore they do not correspond to a state in the CFT. These are
\begin{eqnarray}
\chi_\sigma &\equiv& \left(L_{-2}-\frac{4}{3}L^2_{-1}\right)\sigma \\
\chi_\epsilon &\equiv& \left(L_{-2}-\frac{3}{4}L^2_{-1}\right)\epsilon.
\end{eqnarray}
As a result, all descendants of $\sigma$ and $\epsilon$ under level $2$ are derivative descendants. The lowest descendant that is not a derivative descendant is $L_{-3}\sigma$ in the $\sigma$ tower and $L_{-4}\epsilon$ in the $\epsilon$ tower (and those with $\bar{L}_{-n}$). We shall see later that the $L_{-4}\epsilon$ operator is responsible for the finite-size corrections of the OPE coefficient $C_{\sigma\epsilon\sigma}$ computed on the lattice.

\section{General critical lattice models}
In this section we consider critical lattice models. Given an operator on the lattice, we investigate how to identify it with a sum of CFT operators in the continuum.

\subsection{Fourier modes of multi-site operators}

Recall that the Fourier mode $\mathcal{O}^s$ of a lattice operator $\mathcal{O}$ is defined by
\begin{equation}
\mathcal{O}^s=\sum_{j} \mathcal{O}(j) e^{isx_j2\pi/N}.
\end{equation}
For a one-site operator at site $j$, the position assignment $x_j=j$ appears uncontroversial. However, for an operator $\mathcal{O}(j)$ supported on multiple sites, say from site $j$ to site $j+n$, the position $x_j$ is not uniquely determined. We have to  decide how to assign a specific position $x_j\in (j,j+n)$ to it. Different assignments will lead to different expansions in terms of CFT operators. However, it can be shown that any two such expansions have the same dominant CFT scaling operator, and the difference between the two expansions is dominated by the derivative of this dominant CFT operator.

Let us illustrate the above with an example for the critical Ising model. Consider $\mathcal{O}_1(j)=-X(j)X(j+1)$. We have seen that its CFT expansion $\mathcal{O}^{\CFT}_1$ includes both $\mathbb{1}^{\CFT}$ and $\epsilon^{\CFT}$ contributions. For this lattice operator, we may assign e.g. $x_j=j$, $x_j=j+1/2$, pr $x_j=j+1$. Only the second choice preserves spatial parity, and therefore no $\partial_x \epsilon^{\CFT}$ term is allowed in the expansion of $O^{\CFT}_1$. The other two choices would result in a $\partial_x\epsilon^{\CFT}$ term in the expansion of $\mathcal{O}^{\CFT}_1$, which is in accordance with the fact that our assignment of position has explicitly broken spatial parity. Nevertheless, the expansion coefficients in front of $\mathbb{1}^{\CFT}$ and $\epsilon^{\CFT}$ are independent of our assignment of position $x_j$.

The specific assignment $x_j=j+1/2$ for $\mathcal{O}_1(j)$ is important when combining $\mathcal{O}_1(j)=-X(j)X(j+1)$ with $\mathcal{O}_2(j)=-Z(j)$ to form the Hamiltonian density $h(j)=\mathcal{O}_1(j)+\mathcal{O}_2(j)$. In order for the Fourier mode $h^s$ to correspond to a linear combination of Virasoro generators $L^{\CFT}_{-s}+\bar{L}^{\CFT}_s$, it has been shown numerically \cite{milsted_extraction_2017} that the correct choice is $x_j=j+1/2$ for $O_1(j)$ and $x_j=j$ for $O_2(j)$. If we have chosen a different $x_j$ for $O_1(j)$, the Fourier mode $h^s$ $(s\neq 0)$ would connect states in identity tower and $\epsilon$ tower. This is exactly the consequence of the $\partial_x \epsilon^{\CFT}$ term in the expansion of $h$. 
 
In the main text, for $Z_2$ odd operators we have used a simple \textit{middle point rule} to assign positions, that is, an operator with support from site $j$ to $j+n$ is assigned position $x_j=j+n/2$. $Z_2$ even operators are assigned positions at the middle point of their fermionic representations in Table \ref{table2}.  The fermions $\psi(2j-1)$ and $\psi(2j)$ are assigned positions $x=j-1/4$ and $x=j+1/4$ respectively (according to the picture that one spin degree of freedom splits into two Majorana fermion degree of freedom). For most $Z_2$ even operators that are considered here, it coincides with the middle point rule in the spin representation. The exceptions are: $X(j)Y(j+1)$ is assigned position $x_j=j+3/4$ and $Y(j)X(j+1)$ is assigned position $x_j=j+1/4$. 

We point out that different position assignments can also be chosen which are equally valid, as long as a consistent choice of convention is kept throughout the computations. 

\subsection{Fixing phases of low energy eigenstates}
Diagonalization of the Hamiltonian yields a set of eigenstates with arbitrary complex phases. However, in the CFT calculations needed in our cost function (used in the main text to identify lattice operators with CFT operators), we are comparing lattice matrix elements with CFT matrix elements directly. To do this in a meaningful way, we first have to fix the complex phases of the low energy eigenstates of the critical spin chain using the same conventions used in the CFT. This is achieved by requiring certain matrix elements of lattice operators to have the same phases as in their CFT counterparts. 

First, Fourier modes $h^s \sim L^{\CFT}_{-s}+\bar{L}^{\CFT}_{s}$  $(s\neq0)$ of the lattice Hamiltonian density $h(j)$ are ladder operators in the scaling limit. In a CFT,
\begin{eqnarray}
\langle (L_{-n}\psi_\alpha)^{\CFT}|L^{\CFT}_{-n}|\psi^{\CFT}_\alpha\rangle>0, \\
\langle (\bar{L}_{-n}\psi_\alpha)^{\CFT}|\bar{L}^{\CFT}_{-n}|\psi^{\CFT}_\alpha\rangle>0.
\end{eqnarray}
Accordingly, we will require that the equivalent lattice matrix elements also satisfy
\begin{eqnarray}
\langle L_{-n}\psi_\alpha|h^{s=n}|\psi_\alpha\rangle > 0,\\
\langle \bar{L}_{-n}\psi_\alpha|h^{s=-n}|\psi_\alpha\rangle >0,
\end{eqnarray}
up to finite-size corrections. In practice, for a given conformal tower, we fix relative phases between descendant states and the primary state $|\phi\rangle$ level by level. Starting with $|\psi_\alpha\rangle=|\phi\rangle$, we require the above matrix elements with $n=1$ and $n=2$ to be real and positive. Then we continue to $|\psi_\alpha\rangle=|L_{-n}\phi\rangle$ and $|\bar{L}_{-n}\phi\rangle$ ($n=1,2$) and fix the phases of higher level descendants. This is done until all the selected states in the cost function have their phases fixed with respect to the primary states.

 In the remaining we would like to fix relative phases between primary states $|\phi_\alpha\rangle$. In the CFT,
 \begin{equation}
 \langle\phi^{\CFT}_\alpha|\phi^{\CFT,s_\alpha}_\alpha|0^{\CFT}\rangle>0.
 \end{equation}
 On the lattice, we first find an operator $O$ which has $\phi^{\CFT}_\alpha$ in its expansion, and then require 
\begin{equation}
\langle \phi_\alpha|O^{s_\alpha}|0\rangle
\end{equation}
to be real and positive.

In the Ising CFT, primary fields are Hermitian,
\begin{equation}
\phi^{\CFT\dagger}_\alpha(0,x)=\phi^{\CFT}_\alpha(0,x).
\end{equation}
Therefore, for the critical Ising model, we can choose
\begin{equation}
\langle \sigma|X^{s=0}|0\rangle>0, ~~~~~\langle \epsilon|(XX)^{s=0}|0\rangle>0,
\end{equation}
although we could have chosen other operators (for example, $XZ+ZX$ for $\sigma^{\CFT}$ and $-Z$ for $\epsilon^{\CFT}$, which turns out to be completely equivalent). After that, we have fixed the complex phases in all low energy eigenstates relative to the ground state.

\subsection{OPE coefficients}
Following the CFT expression, OPE coefficients can be computed on the lattice by
\begin{equation}
C_{\alpha\beta\gamma}=\left(\frac{2\pi}{N}\right)^{-\Delta_\beta}\langle\phi_\alpha|\phi_\beta(0)|\phi_\gamma\rangle,
\end{equation}
where $\phi_\beta$ is the lattice operator corresponding to $\phi^{\CFT}_\beta$. Since each state is an eigenstate of the translation operator, the above equation can also be written as
\begin{equation}
C_{\alpha\beta\gamma}=\left(\frac{2\pi}{N}\right)^{-\Delta_\beta}\langle\phi_\alpha|\phi^{s_\alpha-s_\gamma}_\beta|\phi_\gamma\rangle
\end{equation} 
because only the Fourier mode with momentum $s_{\alpha}-s_{\gamma}$ is consistent with momentum conservation.

Note that the exact lattice representation $\phi_\beta$ of a CFT scaling operator $\phi_{\beta}^{\CFT}$ is not expected to be supported on a finite number of lattice sites in general. However, here we always work with a truncated set $\mathcal{A}$ of scaling operators and the resulting lattice operator $\mathcal{O}_{\phi_\beta}$ is only approximate but with finite local support and such that the above matrix elements differ by finite-size corrections. 

After fixing the complex phases of lattice eigenstates as discussed above, for the Ising model we numerically find that
\begin{equation}
C_{\sigma\epsilon\sigma}=\left(\frac{2\pi}{N}\right)^{-\Delta_\epsilon}\bra{\sigma} \mathcal{O}^{s=0}_\epsilon\ket{\sigma}
\end{equation}
and
\begin{equation}
C_{\epsilon\sigma\sigma}=\left(\frac{2\pi}{N}\right)^{-\Delta_\sigma}\bra{\epsilon} \mathcal{O}^{s=0}_\sigma\ket{\sigma}
\end{equation}
are both real and positive for all approximate lattice representations of primary operators that were used.

\subsection{Sources of errors}
In the main text, we have described the cost function that is minimized to obtain the truncated CFT operator expansion $\tilde{\mathcal{O}}^{\CFT}$ corresponding to a lattice operator $\mathcal{O}$. Here we consider possible sources of errors in the construction, and how we can reduce the error in practice. There are three sources of errors.
\begin{enumerate}
\item The CFT operator space is truncated to a finite set $\mathcal{A}$. This precludes an exact correspondence between CFT operators and lattice operators.
\item The lattice has a finite number $N$ of sites, which leads to errors (due to subleading finite-size corrections) in the numerical estimates of the scaling dimensions and central charge used in order to evaluate CFT matrix elements in the cost function.
\item The numerical diagonalization of the Hamiltonian (e.g. using matrix product states MPS) produces approximate eigenstates (e.g. due to the finite bond dimension of the MPS).
\end{enumerate}
It will be argued in this section that the first source causes errors in the expansion coefficients that decay as $1/N^{p}$, with the power depending on the truncated space $\mathcal{A}$ of CFT operators. The power-law convergence of expansion coefficients is later confirmed numerically by the results obtained with the Ising model. We also briefly comment on the other two sources of error, which are assumed to not be the dominant ones.  

\subsubsection{Errors due to a truncated set $\mathcal{A}$ of CFT operators}
Given a truncated set $\mathcal{A}$ of CFT scaling operators $\{\psi^{\CFT}_{\alpha}(x)\}$ and a lattice operator $\mathcal{O}$, we find the coefficients $a_\alpha(N)$ such that
\begin{equation}
\label{Truncate}
\tilde{O}^{\CFT}(x)=\sum_\alpha  a_\alpha(N)\psi^{\CFT}_{\alpha}(x)
\end{equation}
minimizes the cost function
\begin{equation}
f^{O}_N(\{a_{\alpha}\})=\sum_{\beta,s}\left|\langle \psi_\beta|O^s|0\rangle-\langle \psi^{\CFT}_\beta|\tilde{O}^{\CFT,s}|0^{\CFT}\rangle\right|^2.
\end{equation}
The exactly correspondent CFT operator $O^{\CFT}$, which typically involves an infinite sum of scaling operators, satisfies 
\begin{equation}
\label{MatrixElementNoIRR}
\langle \psi_\beta|O^s|0\rangle=\langle \psi^{\CFT}_\beta|O^{\CFT,s}|0^{\CFT}\rangle
\end{equation} 
for any $\beta$ and $s$. The goal is to estimate how far the coefficient $a_\alpha(N)$ that minimizes the cost function is away from $a_\alpha$.

Denote by $\mathcal{A}_c$ the set of scaling operators $\{\psi^{\CFT}_{c, \alpha'}(x)\}$ (c is a label for ''complementary'') that together with $\{\psi^{\CFT}_{\alpha}(x)\}$ form a basis that expands $O^{\CFT}(x)$, then we can expand
\begin{equation}
\label{Oexpansion}
O^{\CFT}(x)=\sum_\alpha a_\alpha \psi^{\CFT}_{\alpha}(x)+ \sum_{\alpha'}  b_{\alpha'} \psi^{\CFT}_{c,\alpha'}(x).
\end{equation}
Denote the difference $\delta a_\alpha(N)=a_\alpha-a_\alpha(N)$. Using Eqs.~\eqref{Truncate}-\eqref{MatrixElementNoIRR}, we can express the cost function solely in terms of CFT matrix elements.
\begin{widetext}
\begin{equation}
\label{cost1_noirr}
f^{O}_N(\{a_\alpha(N)\})=\sum_{\beta}\left|\sum_\alpha \delta a_\alpha(N) \langle\psi^{\CFT}_{\beta}|\psi^{\CFT, s_\beta}_\alpha|0^{\CFT}\rangle+\sum_{\alpha'} b_{\alpha'} \langle\psi^{\CFT}_{\beta}|\psi^{\CFT, s_\beta}_{c,\alpha'}|0^{\CFT}\rangle\right|^2.
\end{equation}
\end{widetext}
For simplicity, we first consider the case where the expansion Eq.~\eqref{Oexpansion} only involves operators in one conformal tower. In the limit of large $N$, the second term scales as $N^{-\Delta_c}$, where $\Delta_c$ denotes the smallest scaling dimension of the operators in $\{\psi^{\CFT}_{c, \alpha'}(x)\}$ that have nonzero coefficient $b_{\alpha'_0}\neq 0$. Moreover, the leading contribution of the second term cannot be completely eliminated by fine tuning $\delta a_\alpha(N)$, since otherwise $\psi^{\CFT}_{c,\alpha'_0}(x)$ would be a linear combination of $\{\psi^{\CFT}_{\alpha}(x)\}$. Therefore, the minimum of the cost function $f^{O}_N$ scales as $N^{-2\Delta_c}$. Minimizing the cost function by fine-tuning $\delta a_\alpha(N)$ thus yields 
\begin{equation}
\label{Conv_Coeff}
\delta a_\alpha(N) \sim N^{-(\Delta_c-\Delta_\alpha)}, 
\end{equation}
where $\Delta_\alpha$ is the scaling dimension of $\psi^{\CFT}_\alpha$. 
In practice, we include in $\mathcal{A}$ all possible operators in $\{\psi^{\CFT}_\alpha(x)\}$ up to scaling dimension $\Delta_{\mathrm{max}}$. Then by definition $\Delta_c>\Delta_{\mathrm{max}}$. Therefore, the error becomes smaller as we include operators in $\mathcal{A}$ with higher scaling dimensions so as to increase  $\Delta_{\mathrm{max}}$. Another way to reduce error is to go to large sizes, if this error is dominant over other sources of error mentioned below.
 
If the expansion Eq.~\eqref{Oexpansion} involves operators in different conformal towers, then in Eq.~\eqref{cost1_noirr} the sum over $\beta$ splits into different conformal towers. For each conformal tower, the sum over $\alpha$ and $\alpha'$ are restricted to the same conformal tower. Following the same arguments, we define $\Delta_c$ for each conformal tower as the smallest scaling dimension in $\{\psi^{\CFT}_{c, \alpha'}(x)\}$ in that conformal tower, and Eq.~\eqref{Conv_Coeff} still holds for operators $\psi^{\CFT}_\alpha$ in that conformal tower.

\subsubsection{Other sources of error}
The CFT matrix elements in the cost functions are computed using scaling dimensions and conformal spins extracted from energies and momenta of the excited states,
\begin{eqnarray}
E_\alpha&=&A+\frac{B}{N}\left(\Delta^{\CFT}_\alpha-\frac{c}{12}\right)+O(N^{-\gamma}), \\
P_\alpha&=&\frac{2\pi}{N} s_\alpha,
\end{eqnarray}
where $A,B$ and $\gamma>1$ are non-universal constants. It then follows that the extracted scaling dimensions $\Delta_\alpha$ and the central charge $c$ have finite-size errors compared with $\Delta^{\CFT}_\alpha$ and $c^{\CFT}$. This error can be reduced by going to large sizes $N$ and through a large $N$ extrapolation. The scaling with $1/N$ depends on $\gamma$, which relies on specific irrelevant perturbations in the lattice Hamiltonian. For the Ising model $\gamma=3$, and we have reached several hundred spins in \cite{Zou_conformal_2018} which makes this error on the order of $10^{-7}$ for $\Delta_\epsilon$ and $\Delta_\sigma$. This can be negligible compared to the truncation error in our range of system sizes $N\leq 48$.
 
There are also errors associated with numerical diagonalization of the Hamiltonian. Here we follow \cite{Zou_conformal_2018} to use periodic uniform matrix product states (puMPS) as the diagonalization method, which was shown to result in energy eigensates with numerical errors that grow as we increase their energy. The fidelity of low energy eigenstates can be improved systematically by increasing the bond dimension of the puMPS. Here for the Ising model we use bond dimensions $D\leq 22$ for $N\leq 48$. Errors in fidelity of the first 23 eigenstates ($\Delta\leq 3+1/8$) are at most on the order of $10^{-6}$.

However, we note that the errors introduced during the diagonalization of the Hamiltonian will result in errors in $a_\alpha(N)$ that grow with both the scaling dimension $\Delta_\alpha$ of the operator and the system size $N$. This is because, in the cost function Eq.~\eqref{cost1_noirr}, the coefficient $\delta a_\alpha(N)$ is multiplied by a matrix element that scales as $N^{-\Delta_\alpha}$. To make this error always smaller than the truncation error, in this paper we only kept up to second level descendants in the set $\mathcal{A}$ and system sizes $N\leq 48$ when analysing the Ising model.

A more in-depth discussion of numerical errors in the Ising model can be found in the last appendix.

\subsubsection{Error in numerical estimates of OPE coefficients}

The OPE coefficients are approximately computed by
\begin{equation}
\label{OPEfinitesize}
C_{\alpha\beta\gamma}=\left(\frac{2\pi}{N}\right)^{-\Delta_\beta}\langle\phi_\alpha|\mathcal{O}^{s_\alpha-s_\gamma}_{\phi_\beta}|\phi_\gamma\rangle,
\end{equation}
where $\mathcal{O}_{\phi_\beta}$ is a lattice operator that corresponds to $\mathcal{O}^{\CFT}_{\phi_\beta}\approx\phi^{\CFT}_\beta$. Expanding in terms of scaling operators,
\begin{equation}
\label{ApproxCFT}
\mathcal{O}^{\CFT}_{\phi_\beta}=a_0 \phi^{\CFT}_\beta+\sum_{\beta'\geq 1} a_{\beta'} \psi^{\CFT}_{\beta'},
\end{equation}
where $a_0\approx 1$ and $\psi^{\CFT}_{\beta'}$ represents other scaling operators. Then
\begin{eqnarray}
\label{OPEerr1}
&C_{\alpha\beta\gamma}&-C^{\CFT}_{\alpha\beta\gamma}=(a_0-1)C^{\CFT}_{\alpha\beta\gamma}\\
\label{OPEerr2}
&+& \left(\frac{2\pi}{N}\right)^{\Delta_\beta}\sum_{\beta'\geq 1} a_{\beta'} \langle\phi^{\CFT}_\alpha|\psi^{\CFT,s_\alpha-s_\gamma}_{\beta'}|\phi^{\CFT}_\gamma\rangle.
\end{eqnarray}
We see that the error of $C_{\alpha_\beta\gamma}$ has two contributions. The first contribution, Eq.~\eqref{OPEerr1}, contributes to a constant proportional to $a_0-1$, which is determined by the accuracy of the expansion coefficients of each lattice operator that are used to construct $\mathcal{O}_{\phi_\beta}$. The second contribution, Eq.~\eqref{OPEerr2}, scales as $N^{-(\Delta_{\beta'_0}-\Delta_\beta)}$, where $\Delta_{\beta'_0}$ is the scaling dimension of $\psi^{\CFT}_{\beta'_0}$ that appears in Eq.~\eqref{OPEerr2}. 

Therefore, in order to increase the accuracy of $C_{\alpha\beta\gamma}$, we can either compute Eq.~\eqref{OPEfinitesize} at larger sizes $N$, or obtain more significant digits of $a_0$. We shall see in the next appendix for the Ising model that both could lead to a significant improvement of accuracy.

\section{The critical Ising model}
In this section we first exactly compute some matrix elements in the low energy spectrum of the Ising model using the free fermion representation. Then we use these exact matrix elements to obtain an exact expression for some of the (numerical) coefficients in Table \ref{table}. Finally, we analyse the numerical results.
 
\subsection{Free fermion representation}
Consider the critical Ising model with periodic boundary conditions,
\begin{eqnarray}
H&=&\sum_{j=1}^N h(j), \\
h(j)&=&-X(j)X(j+1)-Z(j),
\end{eqnarray}
where site $j=N+1$ is identified with site $j=1$. The model has a $Z_2$ symmetry generated by
\begin{equation}
G_S=\prod_{j=1}^N Z_j.
\end{equation}
It is easy to check that $[G_S,H]=0$. $G_S$ and $H$ can be then simultaneously diagonalized, resulting in the eigenvectors of $H$ divided into parity even ($G_S=1$) and parity odd ($G_S=-1$) sectors.

The Jordan-Wigner transformation
\begin{eqnarray}
\psi(2j-1)=\left(\prod_{1\leq k<j} Z(k)\right) \frac{X(j)}{\sqrt{2}} \\
\psi(2j)=\left(\prod_{1\leq k<j} Z(k)\right) \frac{Y(j)}{\sqrt{2}}
\end{eqnarray}
maps the Ising model with $N$ spins to a spinless fermion chain with $2N$ Majorana fermions, where
\begin{equation}
\label{Hermiticity_psi}
\psi^{\dagger}(j)=\psi(j)
\end{equation}
and
\begin{equation}
\label{Commutation_psi}
\{\psi(j),\psi(l)\}=\delta_{jl}.
\end{equation}
Local spin operators with odd $Z_2$ symmetry are mapped to a string of fermion operators, while those with even $Z_2$ symmetry are mapped to local operators in the fermion picture. We list some examples in table \ref{table2}.
\begin{table*}[ht]
\begin{tabular}{|c|c|c|}
\hline
       spin operator & fermion operator  \\ \hline
       $X(j)X(j+1)$     &  $-2i\psi(2j)\psi(2j+1)$   \\ \hline
       $Z(j)$     &   $-2i\psi(2j-1)\psi(2j)$\\ \hline
       $X(j)Y(j+1)$     &   $-2i\psi(2j)\psi(2j+2)$  \\ \hline
       $Y(j)X(j+1)$     &   $2i\psi(2j-1)\psi(2j+1)$   \\ \hline
       $Y(j)Y(j+1)$     &   $2i\psi(2j-1)\psi(2j+2)$      \\ \hline
       $Z(j)Z(j+1)$     &    $-4\psi(2j-1)\psi(2j)\psi(2j+1)\psi(2j+2)$   \\ \hline
       $X(j)I(j+1)X(j+2)$     &  $-4\psi(2j)\psi(2j+1)\psi(2j+2)\psi(2j+3)$   \\ \hline
       $X(j)Z(j+1)X(j+2)$     &  $-2i\psi(2j)\psi(2j+3)$        \\ \hline
\end{tabular}
       \caption{\label{table2} Lattice operators with even $Z_2$ symmetry and their representation using Majorana fermion operators}
\end{table*}

Let us represent the Ising Hamiltonian with the fermionic variables, 
\begin{eqnarray}
H &=&\sum_{j=1}^{N-1} 2i\left[\psi(2j-1)\psi(2j)+\psi(2j)\psi(2j+1)\right] \\
   &+& 2i \left[\psi(2N-1)\psi(2N)-G_S \psi(2N)\psi(1)\right]. 
\end{eqnarray}

One has to be careful with the boundary term. In the even $Z_2$ sector, the fermionic chain has the anti-periodic boundary condition, $\psi(2N+j)=-\psi(j)$, which is usually referred to as the Neveu-Schwarz (NS) sector. On the other hand, in the odd $Z_2$ sector, the fermionic chain has periodic boundary condition, $\psi(2N+j)=\psi(j)$, which is usually referred to as the Ramond (R) sector. We shall only consider the even $Z_2$ sector below. 

Assuming the NS boundary condition, the Hamiltonian can be written more compactly as
\begin{equation}
\label{IsingFermion_e}
H = \sum_{j=1}^{2N} 2i \psi(j)\psi(j+1)
\end{equation}

Note that the Hamiltonian is quadratic in fermionic variables. This makes it a free theory, which can be solved exactly using a Fourier transformation (below).

\subsection{Symmetry and self-duality}
In fermionic variables, the generator of the $Z_2$ symmetry can be expressed as
\begin{equation}
G_S=(-2i)^N \prod_{j=1}^{2N} \psi(j).
\end{equation}
It is then easy to see that $G_S$ commutes with fermionic bilinears, of the form $\psi(j)\psi(j')$, but anti-commutes with operators linear in $\psi(j)$. 

The Ising model at criticality possesses the famous Kramers-Wannier self-duality. This becomes a translation in the Majorana fermion picture,
\begin{equation}
\psi(2j-1)\rightarrow \psi(2j), \,\, \psi(2j)\rightarrow \psi(2j+1).
\end{equation} 
The Hamiltonian Eq.~\eqref{IsingFermion_e} is then manifestly invariant under the duality transformation. Note that applying the duality transformation twice corresponds to a translation by one site in spin variables.  

Using the fermionic representation of local spin operators (Table \ref{table2}), it is easy to see how they transform into each other under the duality transformation.
\begin{eqnarray}
Z(j)&\rightarrow& X(j)X(j+1) \nonumber\\
        &\rightarrow& Z(j+1),  \\
Y(j)X(j+1)&\rightarrow& -X(j)Y(j+1) \nonumber \\
                  &\rightarrow& Y(j+1)X(j+2),  \\
Y(j)Y(j+1)&\rightarrow& -X(j)Z(j+1)X(j+2) \nonumber \\
                     &\rightarrow& Y(j+1)Y(j+2),  \\
Z(j)Z(j+1)&\rightarrow& X(j)I(j+1)X(j+2) \nonumber \\
                    &\rightarrow& Z(j+1)Z(j+2). 
\end{eqnarray}
We can then combine them into duality even operators (e.g., $XX+Z$) and duality odd operators (e.g., $XX-Z$). In the Ising model, operators that are $Z_2$ even and duality even belong to the conformal tower of the identity primary, while operators that are $Z_2$ even and duality odd belong to the $\epsilon$ tower. $Z_2$ odd operators (which are not considered here) belongs to the $\sigma$ tower.

\subsection{Ground state correlation functions}
Define the Fourier modes of fermion operators as
\begin{equation}
\psi(p)=\frac{1}{\sqrt{2N}}\sum_{j=1}^{2N} \psi(j) e^{ipj},
\end{equation}
where the boundary conditions impose
\begin{equation}
p=p_n\equiv \frac{2\pi}{2N}\left(n+\frac{1}{2}\right), \,\, n=-N,-N+1,\cdots,N-1.
\end{equation}
The inverse Fourier transform is 
\begin{equation}
\label{Majorana_Fourier}
\psi(j)=\frac{1}{\sqrt{2N}}\sum_{n=-N}^{N-1} \psi(p_n) e^{-ip_nj},
\end{equation}
Eqs~\eqref{Hermiticity_psi}-\eqref{Commutation_psi} imply 
\begin{eqnarray}
\psi^{\dagger}(p)=\psi(-p), \\
\{\psi(p),\psi(q)\}=\delta_{p+q,0}.
\end{eqnarray}
It follows that $\psi(-p),\psi(p)$ can be understood as fermionic creation and annihilation operator of the mode $p$. Therefore we shall only include the modes with $p>0$ as independent variables.

The Hamiltonian can be rewritten as
\begin{equation}
H=\sum_{n=0}^{N-1}( -2\sin p_n + 4 \sin p_n \psi^{\dagger}(p_n)\psi(p_n)).
\end{equation}
It follows that the ground state satisfies
\begin{equation}
\psi(p_n)|0\rangle=0, \,\, n=0,1,\cdots,N-1.
\end{equation}
The two point correlation function in momentum space is then
\begin{equation}
\label{2pt_p}
\langle 0|\psi(p_n)\psi(-p_m)|0\rangle=\delta_{nm}, 
\end{equation}
for $p_n>0$ and $0$ otherwise.
Fourier transforming back to position space yields
\begin{equation}
\langle 0|\psi(j)\psi(l)|0\rangle=\frac{1}{2N}\sum_{n=0}^{N-1} e^{-ip_n(j-l)}.
\end{equation}
In the thermodynamic limit $N\rightarrow\infty$, the above sum becomes an integral, 
\begin{eqnarray}
\frac{1}{2\pi}\int_0^{\pi} dp\, e^{-i(j-l)p} &=& \frac{i}{2\pi(j-l)}(e^{-i\pi(j-l)}-1) \\
&=& -\frac{i}{\pi(j-l)}, \,\,\, j-l \,\,\,\mathrm{odd},
\end{eqnarray}
and $0$ if $j-l$ is even and nonzero.

For later use, we also present how the correlation function at finite $N$ behaves,
\begin{equation}
\label{2pt_subleading}
\langle0|\psi(j)\psi(l)|0\rangle=-\frac{i}{j-l}\left(1+\frac{\pi^2}{24N^2} (j-l)^2+O(N^{-4})\right),
\end{equation}
if $j-l$ is odd. 

When $j=l$, we have
\begin{equation}
\langle 0|\psi(j)^2|0\rangle=\frac{1}{2}.
\end{equation}
We then have all the two point correlation functions in the thermodynamic limit. 

For example, 
\begin{equation}
\lim_{N\rightarrow\infty}\langle 0|Z(j)|0\rangle=-2i\langle 0|\psi(2j-1)\psi(2j)|0\rangle=\frac{2}{\pi}.
\end{equation}
And similarly,
\begin{eqnarray}
\lim_{N\rightarrow\infty}\langle 0|X(j)X(j+1)|0\rangle &=& \frac{2}{\pi} \\
\lim_{N\rightarrow\infty}\langle 0|Y(j)Y(j+1)|0\rangle &=& -\frac{2}{3\pi} \\
\lim_{N\rightarrow\infty}\langle 0|X(j)Z(j+1)X(j+2)|0\rangle &=& -\frac{2}{3\pi}
\end{eqnarray}

We can then derive higher point correlation functions by Wick's theorem. For example, in the thermodynamic limit,
\begin{eqnarray}
&\langle 0|&\psi(2j)\psi(2j+1)\psi(2j+2)\psi(2j+3)|0\rangle \\
&=&\langle 0|\psi(2j)\psi(2j+1)|0\rangle\langle 0|\psi(2j+2)\psi(2j+3)|0\rangle \nonumber \\
&+&\langle 0|\psi(2j)\psi(2j+3)|0\rangle\langle 0|\psi(2j+1)\psi(2j+2)|0\rangle \nonumber \\
&=&-\frac{4}{3\pi^2}.
\end{eqnarray}
Then
\begin{eqnarray}
\lim_{N\rightarrow\infty}\langle 0|X(j)I(j+1)X(j+2)|0\rangle &=& \frac{16}{3\pi^2} \\
\lim_{N\rightarrow\infty}\langle 0|Z(j)Z(j+1)|0\rangle &=& \frac{16}{3\pi^2} 
\end{eqnarray}
The ground state expectation value of a lattice operator in the thermodynamic limit gives the coefficient of the identity operator in the corresponding CFT operator. For example

\begin{equation}
\label{XX1}
XX \sim \frac{2}{\pi} \mathbb{1}^{\CFT}+\cdots,
\end{equation}
where $\cdots$ represents other scaling operators. In this way, we obtain the coefficient in front of the identity operator for all $Z_2$ even operators in Table \ref{table}, as listed in Table \ref{table3}.

\subsection{Excited states}
Excited states are created by applying creation operators $\psi^{\dagger}(p_n) (p_n>0)$ on the ground state. There are two sets of creation operators at low energy, those with $p_n$ near $p=0$ and near $p=\pi$, corresponding to chiral and anti-chiral excitations. In the $Z_2$ even sector, there is an even number of fermions. The lowest lying excitations are
\begin{eqnarray}
|\epsilon\rangle &=& e^{i\theta_\epsilon}\psi^{\dagger}(p_0)\psi^{\dagger}(p_{N-1})|0\rangle \\
|T\rangle &=& e^{i\theta_T} \psi^{\dagger}(p_0)\psi^{\dagger}(p_1)|0\rangle \\
|\bar{T}\rangle &=& e^{i\theta_{\bar{T}}}\psi^{\dagger}(p_{N-1})\psi^{\dagger}(p_{N-2})|0\rangle.
\end{eqnarray}
The above phases will be determined shortly.

Matrix elements of lattice operators involving these excited states can then be computed by multi-point correlation functions of Majorana operator. For example,
\begin{eqnarray}
&\langle\epsilon|& X(j)X(j+1)|0\rangle \nonumber \\
 &=& -2ie^{-i\theta_\epsilon}\langle 0|\psi(p_{N-1})\psi(p_0)\psi(2j)\psi(2j+1)|0\rangle \nonumber\\
 &=& -2ie^{-i\theta_\epsilon}[\langle 0|\psi(p_0)\psi(2j)|0\rangle\langle 0|\psi(p_{N-1})\psi(2j+1)|0\rangle \nonumber \\
 &-& \langle 0|\psi(p_0)\psi(2j+1)|0\rangle\langle 0|\psi(p_{N-1})\psi(2j)|0\rangle] \nonumber \\
 &=& -2ie^{-i\theta_\epsilon}[\frac{1}{\sqrt{2N}} e^{ip_0 2j} \frac{1}{\sqrt{2N}} e^{ip_{N-1} (2j+1)} \nonumber \\
 &-& \frac{1}{\sqrt{2N}} e^{ip_0 (2j+1)} \frac{1}{\sqrt{2N}} e^{ip_{N-1} 2j}] \nonumber \\
 &=& -\frac{i e^{-i\theta_\epsilon}}{N}(e^{ip_{N-1}}-e^{i p_0}) \nonumber \\
 &=& \frac{2i e^{-i\theta_\epsilon}}{N} \cos{p_0},
 \end{eqnarray}
where the first equality follows from Table \ref{table2}, the second equality follows from the Wick theorem, the third equality follows from Eq.~\eqref{2pt_p} and Eq.~\eqref{Majorana_Fourier}, and the last two equalities follows from $p_0+p_{N-1}=\pi$. 

At large sizes, 
\begin{equation}
\langle\epsilon|X(j)X(j+1)|0\rangle=\frac{2ie^{-i\theta_\epsilon}}{N}\left[1-\frac{\pi^2}{8N^2}+O(N^{-4})\right].
\end{equation}

As stated in Appendix C, we fix the phase $\theta_\epsilon=\pi/2$ by requiring the above matrix element to be real and positive. 

Comparing to the CFT result
\begin{equation}
\langle\epsilon^{\CFT}|\epsilon^{\CFT}(x)|0^{\CFT}\rangle=\frac{2\pi}{L}
\end{equation}
and
\begin{equation}
\langle\epsilon^{\CFT}|\partial^2_\tau\epsilon^{\CFT}(x)|0^{\CFT}\rangle=\left(\frac{2\pi}{L}\right)^3
\end{equation}
(derived from Eq.~\eqref{CFTmatrix}), we can read off
\begin{equation}
\label{XXepsilon}
XX\sim \frac{1}{\pi}\epsilon^{\CFT}-\frac{1}{32\pi}\partial^2_\tau\epsilon^{\CFT}+\cdots,
\end{equation}
where $\cdots$ may contain $\partial^{2}_x\epsilon^{\CFT}$ and higher scaling dimensions in the $\epsilon$ tower, as well as operators in the identity tower.  

Similarly, we can compute
\begin{equation}
\langle T|X(j)X(j+1)|0\rangle=-\frac{i e^{-i\theta_{T}}}{N}e^{i 2\pi s_{T}j/N}(e^{ip_{1}}-e^{i p_0}),
\end{equation}
where $s_T=2$ is the conformal spin of $T$.

Expanding it with respect to $1/N$ gives
\begin{equation}
\langle T|X(j)X(j+1)|0\rangle= \pi\frac{e^{-i\theta_T}}{N^2} e^{i 2\pi s_{T}j/N}.
\end{equation}
By requiring it to be negative, we fix $\theta_T=\pi$. Then we can compare it to
\begin{equation}
\langle T^{\CFT}|T^{\CFT}(x)|0^{\CFT}\rangle= \left(\frac{2\pi}{L}\right)^2\sqrt{\frac{c}{2}} e^{i 2\pi s_{T}x/L},
\end{equation}
where $c=1/2$ is the central charge,
to obtain
\begin{equation}
\label{XXT}
XX\sim -\frac{1}{2\pi} T^{\CFT}+\cdots,
\end{equation}
where $\cdots$ contains other scaling operators. In the same way,
\begin{equation}
\label{XXTbar}
XX\sim -\frac{1}{2\pi} \bar{T}^{\CFT}+\cdots.
\end{equation}
Now we can combine Eqs.~\eqref{XX1},\eqref{XXepsilon},\eqref{XXT},\eqref{XXTbar} to obtain
\begin{equation}
XX\sim \frac{2}{\pi}\mathbb{1}^{\CFT}-\frac{1}{2\pi}(T^{\CFT}+\bar{T}^{\CFT})+\frac{1}{\pi}\epsilon^{\CFT}-\frac{1}{32\pi}\partial^2_\tau\epsilon^{\CFT}+\cdots,
\end{equation}
where $\cdots$ contains other scaling operators with scaling dimension 3 or higher.

Since $XX$ and $Z$ are related by a duality transformation, they have the same coefficients in front of operators in the identity tower, and opposite coefficients in front of operators in the $\epsilon$ tower. Then 
\begin{equation}
Z\sim \frac{2}{\pi}\mathbb{1}^{\CFT}-\frac{1}{2\pi}(T^{\CFT}+\bar{T}^{\CFT})-\frac{1}{\pi}\epsilon^{\CFT}+\frac{1}{32\pi}\partial^2_\tau\epsilon^{\CFT}+\cdots.
\end{equation}

Proceeding as above for other lattice operators as listed in Table \ref{table2}, we reproduce part of Table \ref{table} analytically, as listed in Table \ref{table3}. We note that the subleading term in Eq.~\eqref{2pt_subleading} is important in deriving the coefficient in front of $\partial^2_\tau\epsilon^{\CFT}$ for $XIX$ and $ZZ$, which are quartic in fermionic variables. The coefficient in front of $\partial^2_x\epsilon^{\CFT}$ cannot be computed by the matrix elements $\langle \epsilon|O(j)|0\rangle$ because $\langle \epsilon^{\CFT}|\partial^2_x\epsilon^{\CFT}|0^{\CFT}\rangle=0$. Instead, we have to use matrix elements involving a state in the $\epsilon$ tower with non-vanishing conformal spin, such as $|\partial\epsilon\rangle$.

In Table \ref{table3}, we also show the first $5$ digits of each analytically computed coefficient, to be compared with numerical results in the main text.

\begin{table*}[ht]
 
 \begin{tabular}{|c|c|}
\hline
       Lattice operator  & CFT operator  \\ \hline
       $Z$     &    $(2/\pi) \mathbb{1}-1/(2\pi)(T+\bar{T})-(1/\pi)\epsilon+1/(32\pi)\partial^2_\tau\epsilon$   \\ \hline
       $XX$     &   $(2/\pi) \mathbb{1}-1/(2\pi)(T+\bar{T})+(1/\pi)\epsilon-1/(32\pi)\partial^2_\tau\epsilon$ \\ \hline
       $YY$      &   $-2/(3\pi) \mathbb{1}+3/(2\pi)(T+\bar{T})+(1/\pi)\epsilon-9/(32\pi)\partial^2_\tau\epsilon$   \\ \hline
       $ZZ$      & $16/(3\pi^2) \mathbb{1}-16/(3\pi^2)(T+\bar{T})-16/(3\pi^2)\epsilon+2/(3\pi^2)\partial^2_\tau\epsilon$   \\ \hline
       $XZX$      & $2/(3\pi) \mathbb{1}-3/(2\pi)(T+\bar{T})+(1/\pi)\epsilon-9/(32\pi)\partial^2_\tau\epsilon$       \\ \hline
       $XIX$    &    $16/(3\pi^2) \mathbb{1}-16/(3\pi^2)(T+\bar{T})+16/(3\pi^2)\epsilon-2/(3\pi^2)\partial^2_\tau\epsilon$       \\ \hline
       $-i(XY+YX)$     &    $(1/\pi)\partial_\tau \epsilon$    \\ \hline
       $XY-YX$     &   $(2/\pi)(T-\bar{T})$        \\ \hline
\end{tabular}
 
 \begin{tabular}{|c|c|}
\hline
       Lattice operator  & CFT operator  \\ \hline
       $Z$     &    $0.63662 \mathbb{1}-0.15915(T+\bar{T})-0.31831\epsilon+0.00995\partial^2_\tau\epsilon$   \\ \hline
       $XX$     &   $0.63662 \mathbb{1}-0.15915(T+\bar{T})+0.31831\epsilon-0.00995\partial^2_\tau\epsilon$ \\ \hline
       $YY$      &   $-0.21221 \mathbb{1}+0.47746(T+\bar{T})+0.31831\epsilon-0.08952\partial^2_\tau\epsilon$   \\ \hline
       $ZZ$      & $0.54038 \mathbb{1}-0.54038(T+\bar{T})-0.54038\epsilon+0.06755\partial^2_\tau\epsilon$   \\ \hline
       $XZX$      & $0.21221 \mathbb{1}-0.47746(T+\bar{T})+0.15915\epsilon-0.08952\partial^2_\tau\epsilon$       \\ \hline
       $XIX$    &    $0.54038 \mathbb{1}-0.54038(T+\bar{T})+0.54038\epsilon-0.06755\partial^2_\tau\epsilon$       \\ \hline
       $-i(XY+YX)$     &    $0.31831\partial_\tau \epsilon$    \\ \hline
       $XY-YX$     &   $0.63662(T-\bar{T})$        \\ \hline
\end{tabular}
\caption{\label{table3} Correspondence between lattice operators and CFT operators for the Ising model. The truncated set of CFT operators contains $\mathbb{1},T,\bar{T},\epsilon,\partial_\tau \epsilon, \partial^2_\tau \epsilon$. Coefficients are obtained analytically in the top table. The bottom table is the same as the top table except that coefficients are shown their approximate values to 5 digits to compare with numerical results. The subscript ''CFT'' is omitted in the column of CFT operators. }
\end{table*}


Comparing Table \ref{table3} with Table \ref{table} in the main text, we see that in general, the coefficients in front of CFT operators with lower scaling dimensions are, as expected, more accurate.

We note in passing that, in order to reproduce the expansion for $Z_2$ odd operators analytically, we have to work with states in the Ramond sector and string operators in the fermion language. This is more complicated and we omit it here.

\subsection{Analysis of numerical computations}

Next we discuss the extrapolation to large system size $N$ of the Ising model of the numerical estimates for both the expansion coefficients and the OPE coefficients. 

\subsubsection{Convergence of expansion coefficients for the Ising model} 
We start with the numerical extrapolation of some of the expansion coefficients $a_\alpha$ presented in Table \ref{table} in the main text. We show that the convergence of $a_\alpha(N)$ is as $1/N^p$, where $p=\Delta_c-\Delta_\alpha$ is the predicted power law in Eq.~\eqref{Conv_Coeff}.

The first example is $\mathcal{O}=XZ+ZX$ with 
\begin{eqnarray}
\tilde{\mathcal{O}}^{\CFT}=a_{\sigma}\sigma^{\CFT}+a_{\partial_\tau\sigma}\partial_\tau\sigma^{\CFT} \nonumber \\
\label{XZexpansion}
+a_{\partial^2_x\sigma}\partial^2_x\sigma^{\CFT}+a_{\partial^2_\tau\sigma}\partial^2_\tau\sigma^{\CFT}.
\end{eqnarray}
The error due to using a truncated set $\mathcal{A}$ of CFT scaling operators is determined by $\Delta_c> 2+1/8$. It turns out that this particular operator does not have a contribution from the $\sigma$ tower at level $3$, and therefore $\Delta_c=4+1/8$. Therefore,
\begin{eqnarray}
\delta a_{\sigma}&\sim& N^{-4}, \\
\delta a_{\partial_\tau\sigma}&\sim& N^{-3}, \\
\delta a_{\partial^2_\tau\sigma}&\sim& N^{-2}, \\
\delta a_{\partial^2_x\sigma}&\sim& N^{-2}.
\end{eqnarray}

\begin{figure}[htbp]
  \includegraphics[width=\linewidth]{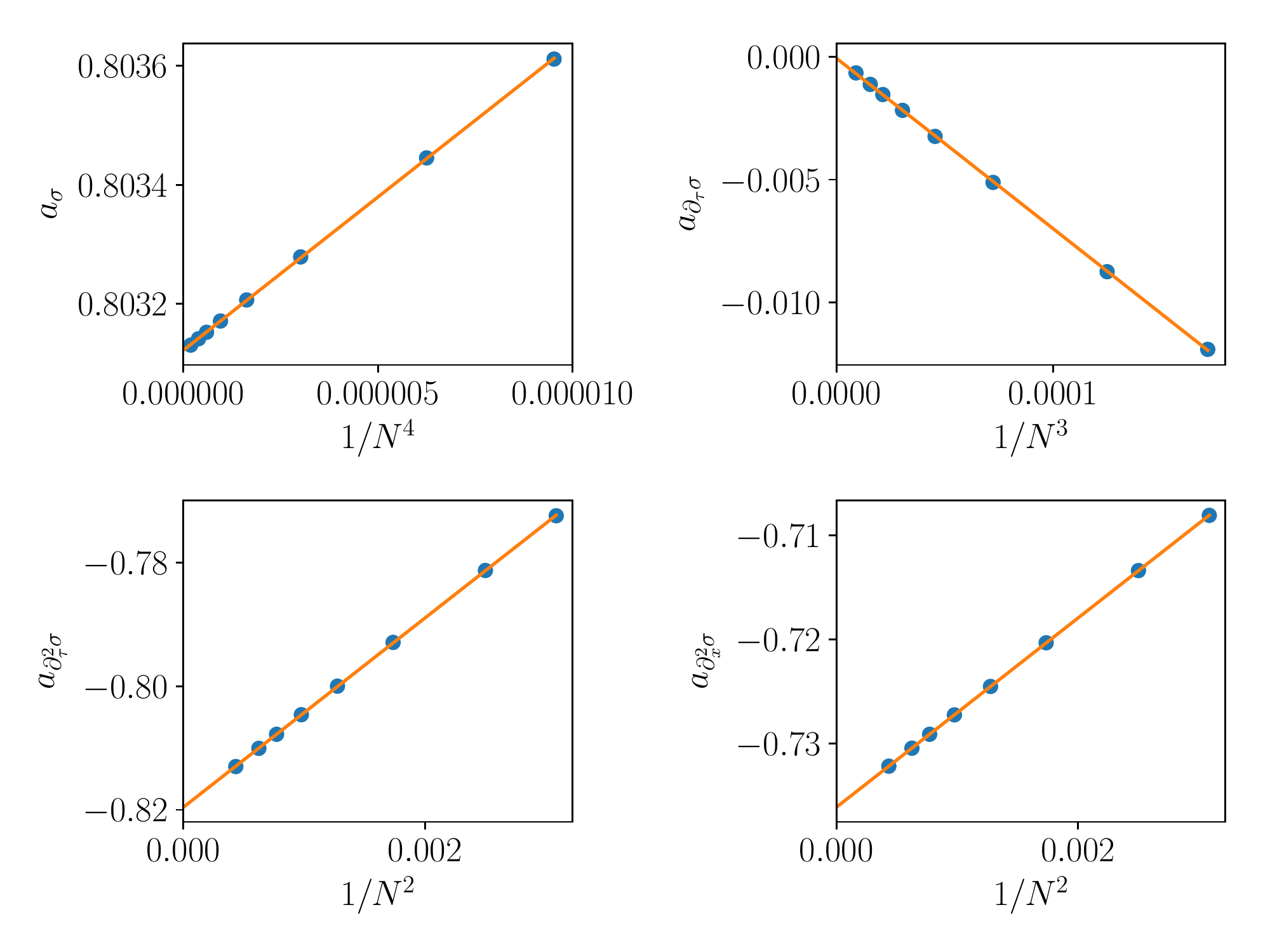}
  \caption{Convergence of the coefficients with Eq.~\eqref{XZexpansion} for $\mathcal{O}=XZ+ZX$. Coefficients are obtained by minimizing the cost function for systems sizes $18\leq N\leq 48$.}
  \label{XZConv}
\end{figure}

The extrapolation of the numerical results is shown in Fig.~\ref{XZConv}.

\begin{figure}[htbp]
  \includegraphics[width=\linewidth]{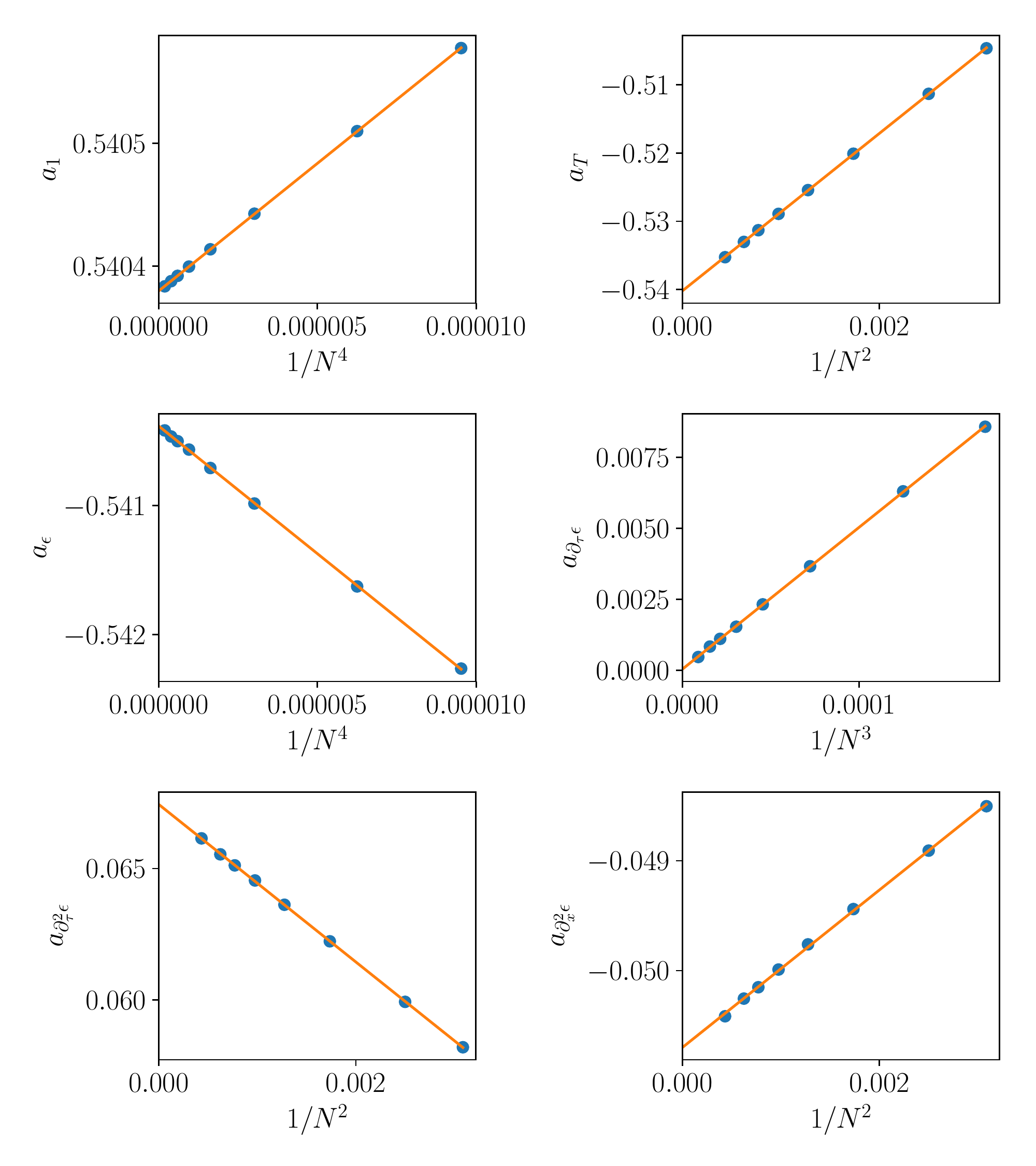}
  \caption{Convergence of the coefficients with Eq.~\eqref{ZZexpansion} for $\mathcal{O}=ZZ$. Coefficients are obtained by minimizing the cost function for systems sizes $18\leq N\leq 48$.}
  \label{ZZConv}
\end{figure}

The second example is $\mathcal{O}=ZZ$ and
\begin{eqnarray}
\tilde{\mathcal{O}}^{\CFT}&=&a_{\mathbb{1}}\mathbb{1}^{\CFT}+a_{T}(T^{\CFT}+\bar{T}^{\CFT}) \nonumber \\
&+&a_{\epsilon}\epsilon^{\CFT}+a_{\partial_\tau\epsilon}\partial_\tau\epsilon^{\CFT} \nonumber \\
\label{ZZexpansion}
&+&a_{\partial^2_x\epsilon}\partial^2_x\epsilon^{\CFT}+a_{\partial^2_\tau\epsilon}\partial^2_\tau\epsilon^{\CFT}.
\end{eqnarray}
In this case, the  leading complementary operator has scaling dimension $\Delta_c=4$ for the identity tower and $\Delta_c=5$ for the $\epsilon$ tower. Then
\begin{eqnarray}
\delta a_{\mathbb{1}}&\sim& N^{-4}, \\
\delta a_{T}&\sim& N^{-2}, \\
\delta a_{\epsilon}&\sim& N^{-4}, \\
\delta a_{\partial_\tau \epsilon}&\sim& N^{-3}, \\
\delta a_{\partial^2_\tau\epsilon}&\sim& N^{-2}, \\
\delta a_{\partial^2_x\epsilon}&\sim& N^{-2}.
\end{eqnarray}
The numerical results are shown in Fig.~\ref{ZZConv}.

We have some additional comments on the extrapolation of the expansion coefficients for lattice operators. First, we cannot determine $\Delta_c$ a priori in general. Instead, we have to try extrapolation using different possible $\Delta_c$ to make the best fit. 

\begin{figure}[htbp]
  \includegraphics[width=\linewidth]{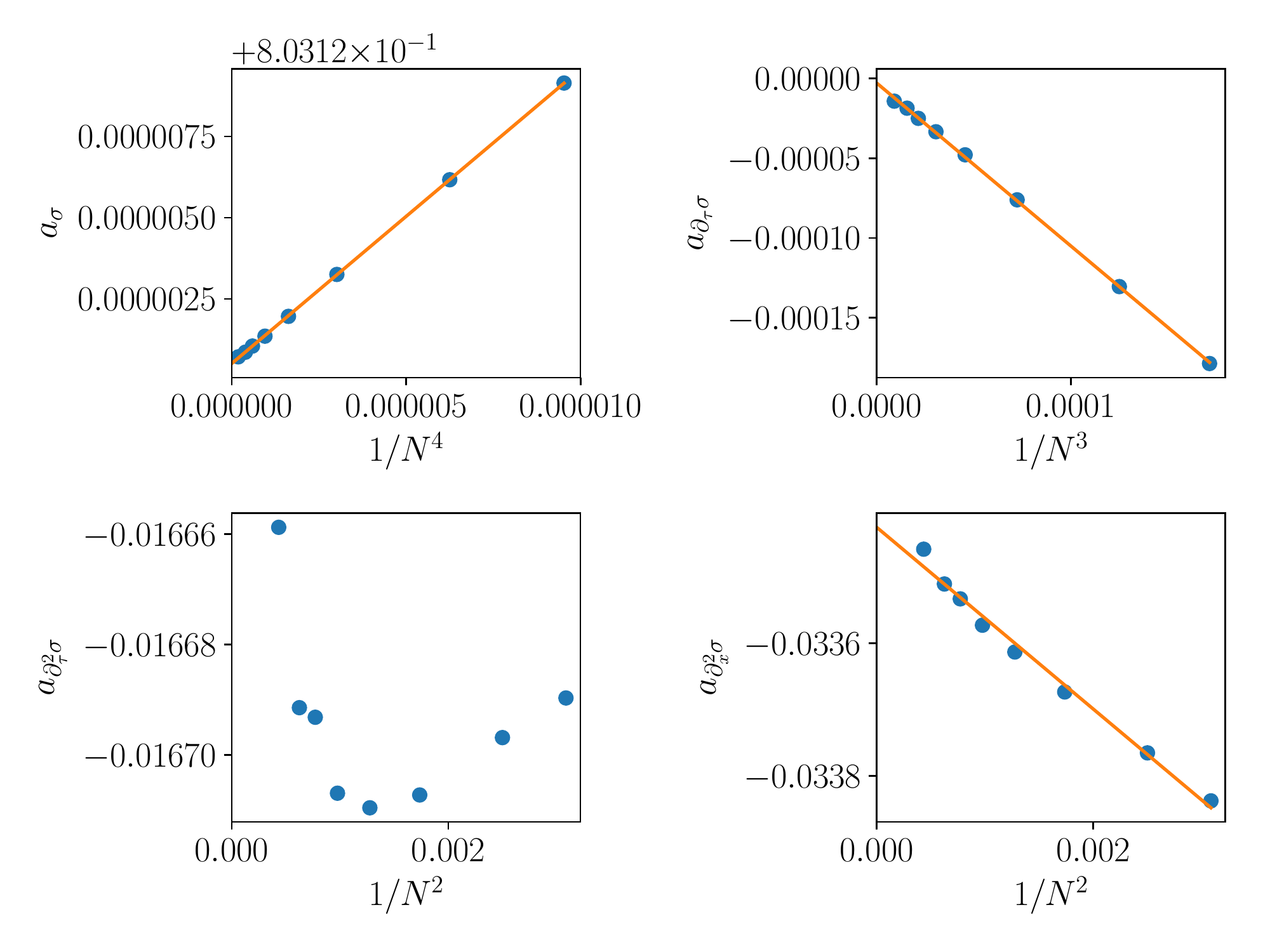}
  \caption{Convergence of the coefficients with Eq.~\eqref{XZexpansion} for $\mathcal{O}=X$. Coefficients are obtained by minimizing the cost function for systems sizes $18\leq N\leq 48$.}
  \label{XConv}
\end{figure}

Second, for some coefficient $a_\alpha(N)$, the error in numerical diagonalization may be important in the extrapolation. For example, this happens for $a_{\partial^2_\tau\sigma}\approx -0.017$ for $\mathcal{O}=X$ with Eq.~\eqref{XZexpansion}, see Fig.~\ref{XConv}.

\begin{figure}[htbp]
  \includegraphics[width=\linewidth]{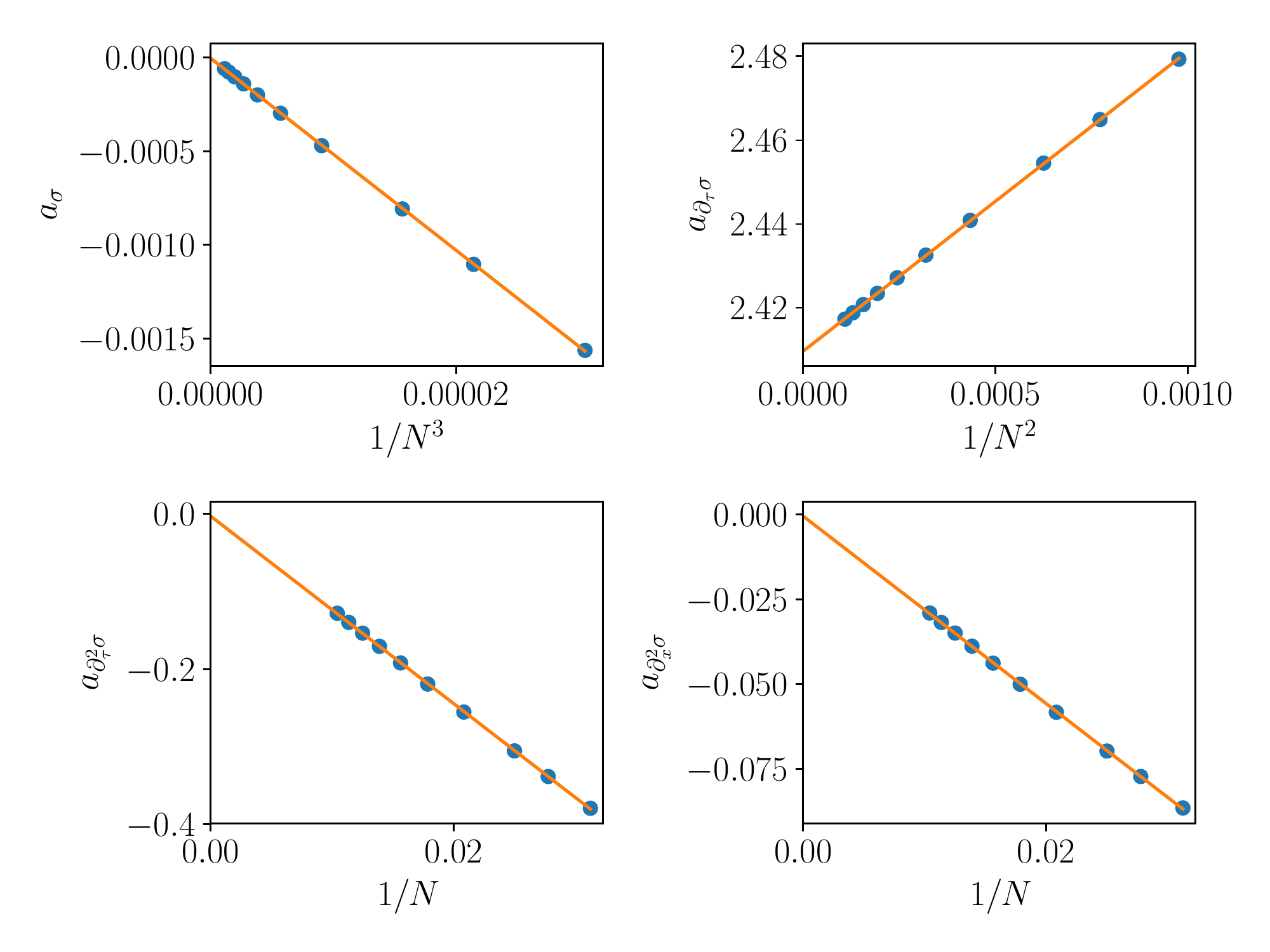}
  \caption{Convergence of the coefficients with Eq.~\eqref{XZexpansion} for $\mathcal{O}=-i(YZ+ZY)$. Coefficients are obtained by minimizing the cost function for systems sizes $36\leq N\leq 96$.}
  \label{YZConv}
\end{figure}

Third, the extrapolation assumes the asymptotic scaling of $\delta a_\alpha(N)$ at large sizes. Numerically, we can determine $a_\alpha$ more accurately by using data from larger sizes, if other sources of error are negligible. In the operators that are considered, we find that for $\mathcal{O}=-i(YZ+ZY)$ with Eq.~\eqref{XZexpansion}, the coefficients $a_{\partial^2_\tau\sigma}$ and $a_{\partial^2_x\sigma}$ are only obviously below $10^{-2}$ when extrapolating with data up to $N=96$, see Fig.~\ref{YZConv}.

\subsubsection{OPE coefficients from the Ising model}
According to Table \ref{table} in the main text, we can three different ansatz for $\mathcal{O}_\sigma$, which all have their corresponding CFT operator in the form of Eq.~\eqref{ApproxCFT}, with $\phi^{\CFT}_\beta=\sigma^{\CFT},a_0\approx 1$. They are
\begin{eqnarray}
\mathcal{O}_{\sigma1}&=&\mu_1 X,  \\
\mathcal{O}_{\sigma2}&=&\mu_2 (XZ+ZX), \\
\mathcal{O}_{\sigma3}&=&\mu_3 (X+\nu_3(XZ+ZX)).
\end{eqnarray}
We quote the expansion coefficients that are used here for reader's convenience,
\begin{eqnarray}
\label{Xexp_num}
X&\sim&0.803121\sigma^{\CFT}-0.017\partial^2_\tau \sigma^{\CFT}+\cdots \\
\label{XZexp_num}
XZ+ZX&\sim&0.803121\sigma^{\CFT}-0.820\partial^2_\tau \sigma^{\CFT}+\cdots,
\end{eqnarray}
where we omit the $\partial^2_x \sigma^{\CFT}$ term because it does not contribute to the OPE coefficient. 
In the following, we shall regard the coefficient of $\sigma^{\CFT}$ in the above two expansions numerically the same, as they coincide with the highest accuracy (6 digits) among all coefficients that are computed.

In order to have $a_0\approx 1$, we determine 
\begin{equation}
\mu_1 = \mu_2 = \mu_3/(1+\nu_3) \approx 1/0.803121\approx 1.24514.
\end{equation}
Since $a_0$ has 6 significant digits and its error can be negligible to the finite-size errors below, we shall ignore the difference between $a_0$ and $1$.

\begin{figure}[htbp]
  \includegraphics[width=\linewidth]{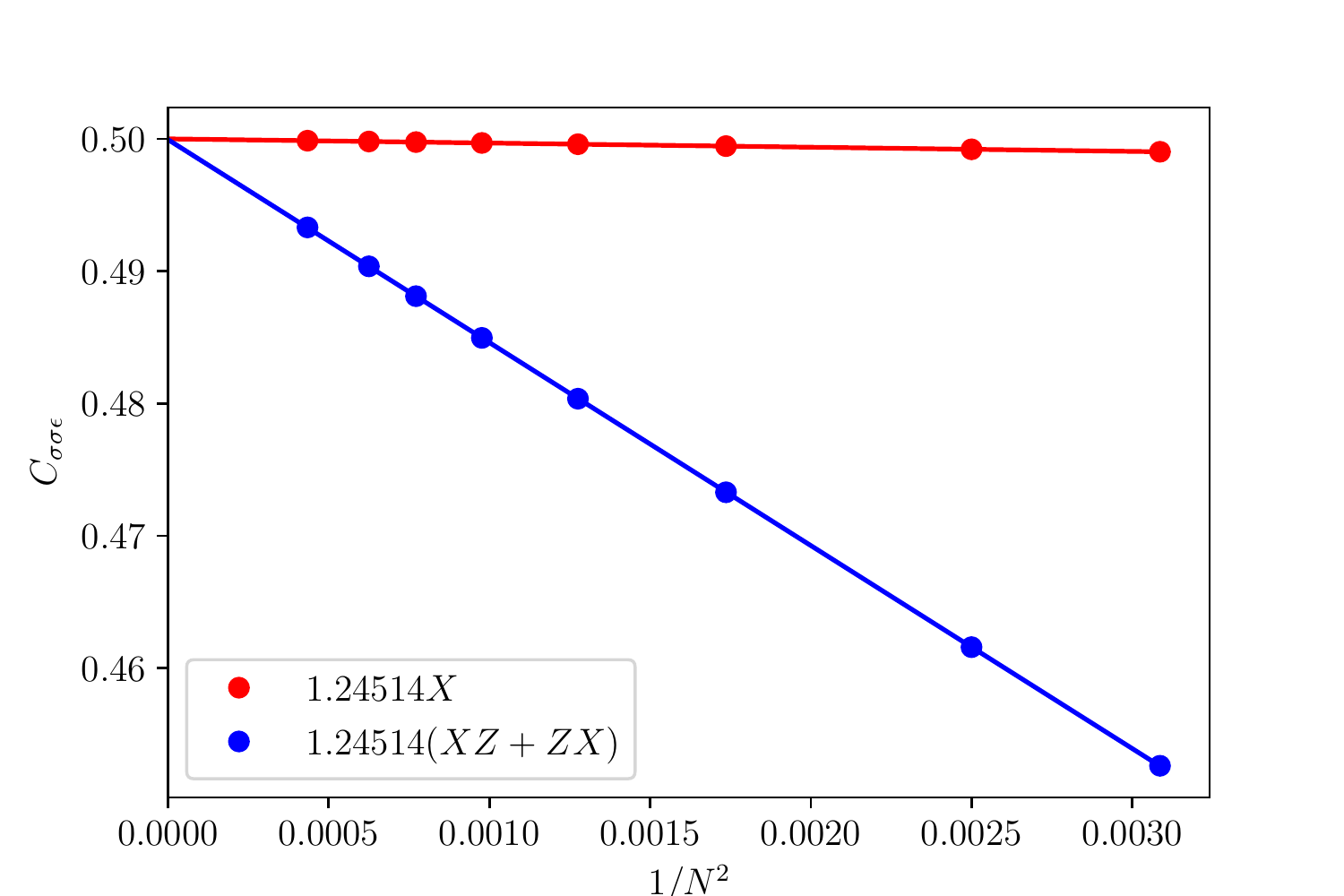}
  \caption{Convergence of the OPE coefficients $C_{\sigma\sigma\epsilon}$ with $\sigma^{\CFT}\sim \mathcal{O}_{\sigma1},\mathcal{O}_{\sigma2}$. Sizes $18\leq N\leq 48$ are used. Linear extrapolation with $1/N^2$ is used. The intercept the the extrapolation are approximately $0.5000003$ and $0.49994$ respectively. The slope are approximately $-0.315$ and $-15.34$ respectively.}
  \label{OPE1}
\end{figure}

The subleading operator in Eq.~\eqref{ApproxCFT} for $\mathcal{O}_{\sigma1}$ and $\mathcal{O}_{\sigma2}$ is $\partial^2_\tau\sigma^{\CFT}$, with coefficient $a_{(\partial^2_\tau\sigma)1}=-0.017\mu_1\approx -0.021$ and $a_{(\partial^2_\tau\sigma)2}=-0.820\mu_2\approx -1.02$. Therefore, the corresponding OPE coefficients are, according to Eqs.~\eqref{OPEerr1}-\eqref{OPEerr2},
\begin{eqnarray}
\label{OPE1_Conv1}
C_{\sigma\sigma\epsilon1}&\approx& C^{\CFT}_{\sigma\sigma\epsilon}+\left(\frac{2\pi}{N}\right)^{2} a_{(\partial^2_\tau\sigma)1} C^{\CFT}_{\sigma\partial^2\sigma\epsilon}  \\
\label{OPE1_Conv2}
C_{\sigma\sigma\epsilon2}&\approx&C^{\CFT}_{\sigma\sigma\epsilon}+\left(\frac{2\pi}{N}\right)^{2} a_{(\partial^2_\tau\sigma)2} C^{\CFT}_{\sigma\partial^2\sigma\epsilon},
\end{eqnarray}
where we have defined
\begin{eqnarray}
C^{\CFT}_{\sigma\partial^2\sigma\epsilon}&=&\left(\frac{2\pi}{L}\right)^{-(2+\Delta^{\CFT}_\sigma)}\langle \sigma^{\CFT}|\partial^2_\tau\sigma^{\CFT}(0)|\epsilon^{\CFT}\rangle \\
&=& \frac{49}{128},
\end{eqnarray}
and where we have used Eq.~\eqref{taudescendantmatrix}. Substituting numbers into Eqs.~\eqref{OPE1_Conv1}-\eqref{OPE1_Conv2} gives
\begin{eqnarray}
\label{OPE1_Conv1_num}
C_{\sigma\sigma\epsilon1}&\approx& 0.5-\frac{0.32}{N^2}  \\
\label{OPE1_Conv2_num}
C_{\sigma\sigma\epsilon2}&\approx&0.5-\frac{15.4}{N^2},
\end{eqnarray}
Eqs.~\eqref{OPE1_Conv1_num}-\eqref{OPE1_Conv2_num} are confirmed with numerical results, see Fig.~\ref{OPE1}.

\begin{figure}[htbp]
  \includegraphics[width=\linewidth]{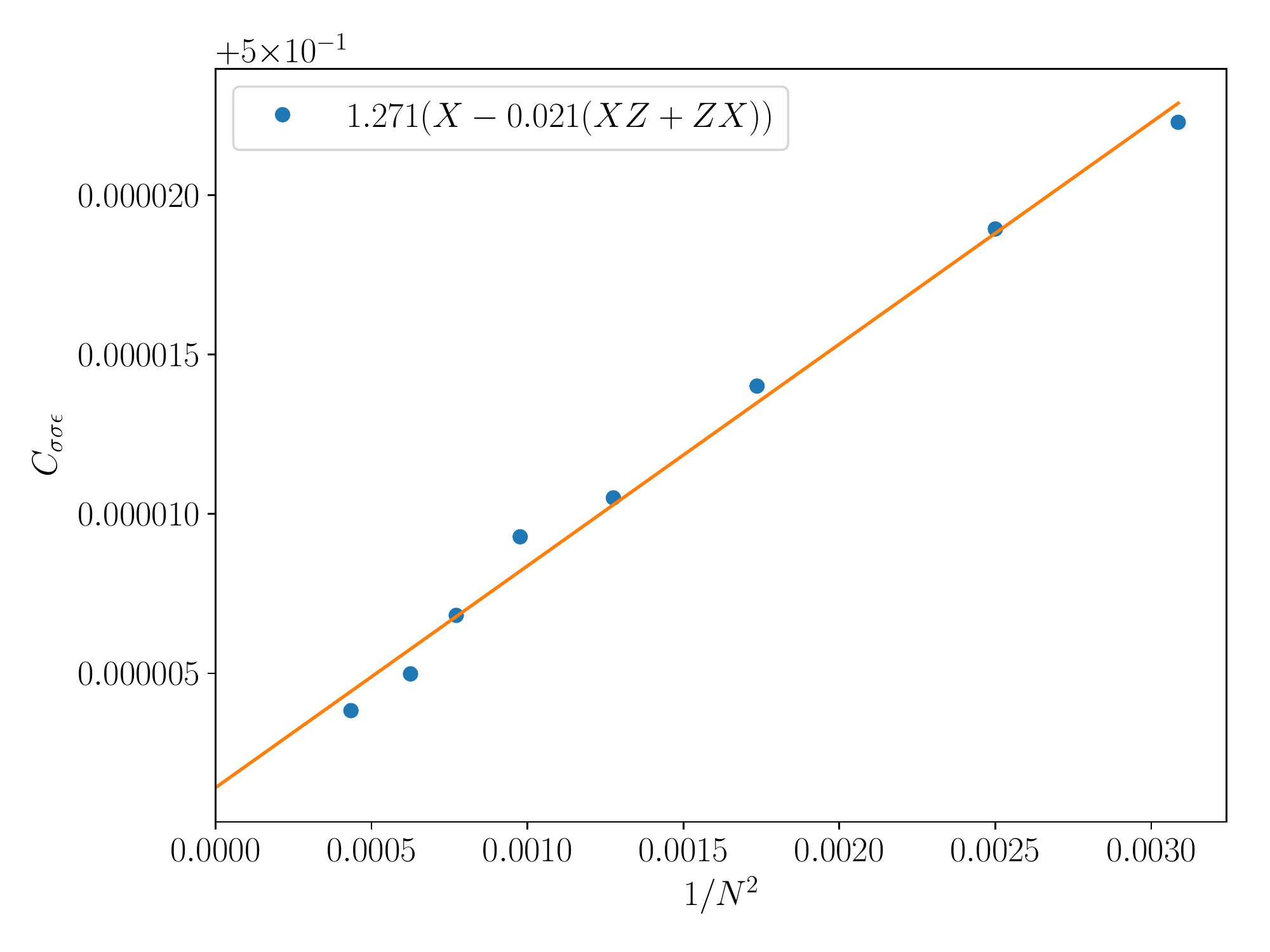}
  \caption{Convergence of the OPE coefficients $C_{\sigma\sigma\epsilon}$ with $\sigma^{\CFT}\sim \mathcal{O}_{\sigma 3}$. The same extrapolations as Fig.~\ref{OPE1} are used. The intercept and slope of the extrapolation are approximately $0.500001$ and $0.003$ respectively.}
  \label{OPE3b}
\end{figure}

We see that using $\sigma^{\CFT}\sim \mathcal{O}_{\sigma1}$ results in much smaller errors, although they scale with the same power of $N$ as the errors in $\sigma^{\CFT}\sim \mathcal{O}_{\sigma2}$. This, of course, originates from the fact that the coefficient of $\partial^2_\tau \sigma^{\CFT}$ in Eq.~\eqref{Xexp_num} has a much smaller amplitude than that in Eq.~\eqref{XZexp_num}. The point of introducing the third lattice realization $\mathcal{O}_{\sigma3}$ is to completely eliminate this contribution. Therefore, $\nu_3\approx -0.017/0.820\approx -0.021$ and it follows that $\mu_3\approx 1.27$. Ideally, this would result in
\begin{equation}
C_{\sigma\sigma\epsilon3}=C^{\CFT}_{\sigma\sigma\epsilon}+O(1/N^4).
\end{equation}
However, since $\nu_3$ has not been determined with enough accuracy, the effect $\partial^2_\tau\sigma$ is not completely removed, and we did not observe a $1/N^4$ scaling. Instead, we find a $1/N^2$ scaling with an almost vanishing coefficient, see Fig.~\ref{OPE3b}.

\begin{figure}[htbp]
  \includegraphics[width=\linewidth]{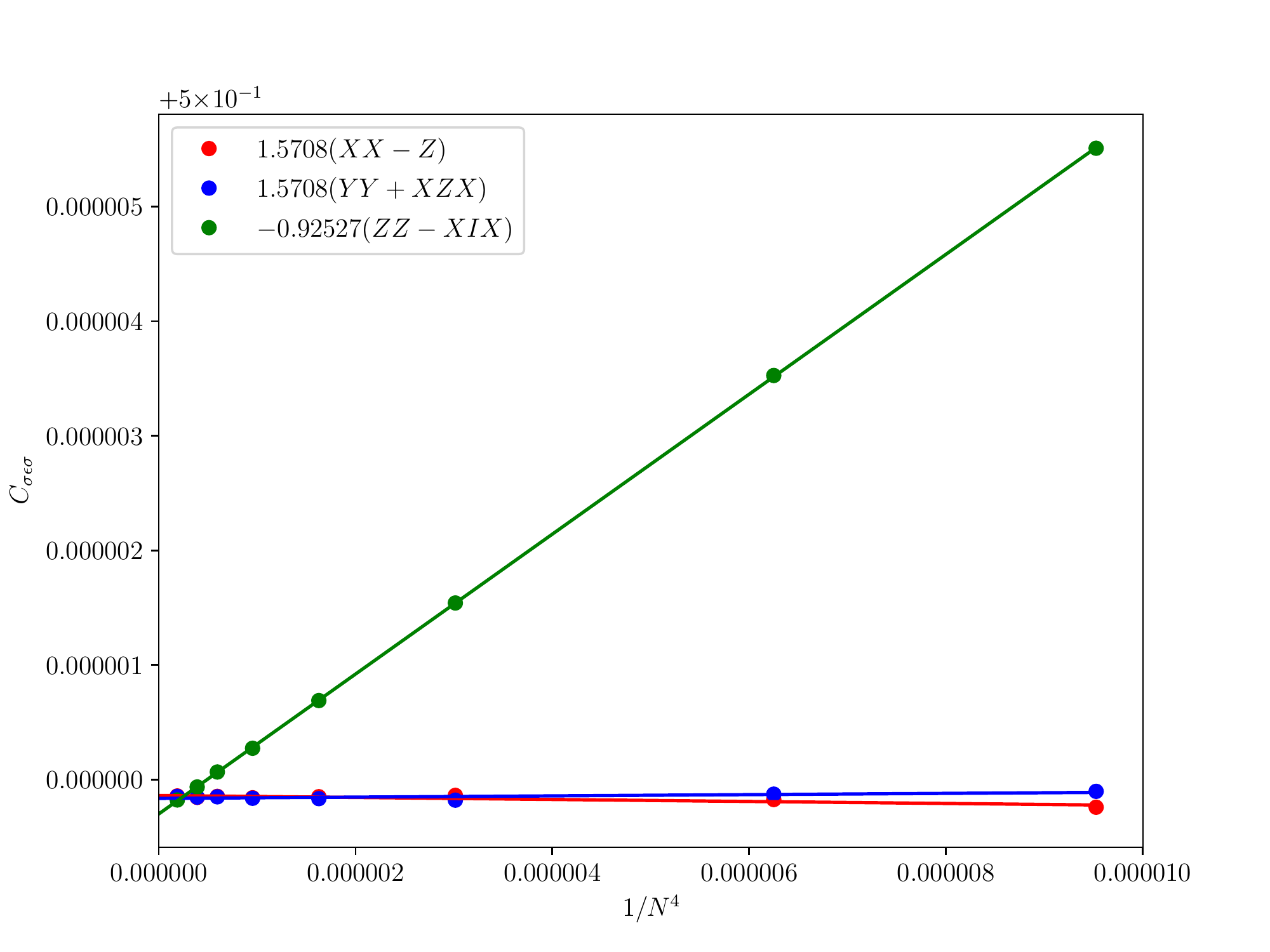}
  \caption{Convergence of the OPE coefficient $C_{\sigma\epsilon\sigma}$ with $\sigma^{\CFT}\sim \mathcal{O}_{\epsilon1}, \mathcal{O}_{\epsilon2}, \mathcal{O}_{\epsilon3}$. Sizes $18\leq N\leq 48$ are used. Linear extrapolation with $1/N^4$ is used. The intercept of the extrapolations are approximately $0.4999999$, $0.4999999$ and $0.4999997$ respectively. Only $\mathcal{O}_{\epsilon3}$ has significant finite-size error in this OPE coefficient, with a slope approximately $0.61$ in the extrapolation.}
  \label{OPE2}
\end{figure}

We can analyze the cases for the other OPE coefficient $C_{\sigma\epsilon\sigma}$ in the same way. We have used 4 different lattice realizations that approximately correspond to $\epsilon^{\CFT}$,
\begin{eqnarray}
\mathcal{O}_{\epsilon 1}&=&\mu_1(XX-Z) \\
\mathcal{O}_{\epsilon 2}&=&\mu_2(YY+XZX) \\
\mathcal{O}_{\epsilon 3}&=&\mu_3(ZZ-XIX) \\
\mathcal{O}_{\epsilon 4}&=&\mu_4(XX-Z+\nu_4(YY+XZX)) ,
\end{eqnarray}
where $\mu_1\approx\mu_2=1/0.63662\approx 1.5708$ and $\mu_3= 1/(-1.0876)\approx -0.92527$. The last operator is to eliminate the $\partial^2_\tau\epsilon^{\CFT}$ contribution. Therefore, $\nu_4=-0.010/0.089\approx -0.11$ and $\mu_4\approx \mu_1/(1+\nu_4)\approx 1.76$. However, since 
\begin{equation}
\langle \sigma^{\CFT}|\partial^2_\tau\epsilon^{\CFT}(0)|\sigma^{\CFT}\rangle=0,
\end{equation}
it does not improve the scaling of $C_{\sigma\epsilon\sigma}$. Therefore, we shall not show the OPE coefficient computed with $\mathcal{O}_{\epsilon4}$, which almost coincides with that with $\mathcal{O}_{\epsilon1}$. The numerical results are shown in Fig.~\ref{OPE2}
We see that only $\mathcal{O}_{\epsilon 3}$ has significant finite-size error in this OPE coefficient, which hints that in the expansion of $\mathcal{O}^{\CFT}_{\epsilon 3}$ there is a significant contribution from $(L_{-4}+\bar{L}_{-4})\epsilon^{\CFT}$. For the other two operators $\mathcal{O}_{\epsilon 1}$ and $\mathcal{O}_{\epsilon 2}$ , the above extrapolation suggests that the only subleading CFT operators are derivative descendants, which do not contribute to the OPE coefficient. This may be explained by the fact that they are quadratic in fermionic variables and the Ising model is a free fermion.

\end{document}